\documentclass[a4paper,11pt]{article}

\usepackage[T1]{fontenc}
\usepackage[a4paper,top=25mm,bottom=25mm,left=28mm,right=28mm]{geometry}

\usepackage{amsmath,amssymb,amsthm,mathtools}
\usepackage{graphicx}
\usepackage[dvipsnames]{xcolor}
\usepackage[export]{adjustbox}
\usepackage{booktabs}
\usepackage{tikz}
\usepackage[normalem]{ulem}

\usepackage[
    colorlinks,
    bookmarks=false,
    citecolor=NavyBlue,
    linkcolor=blue,
    urlcolor=blue
]{hyperref}
\usepackage{doi}
\usepackage[nottoc]{tocbibind}
\usepackage{comment}

\numberwithin{equation}{section}

\newcommand\be{\begin{equation}}
\newcommand\ee{\end{equation}}
\newcommand\bea{\begin{align}}
\newcommand\eea{\end{align}}

\newcommand{\p}{\partial}
\newcommand{\dr}{^{\mathrm{dr}}}

\makeatletter

\def\XXint#1#2#3{%
  \setbox0=\hbox{$#1{#2#3}{\int}$}%
  \vcenter{\hbox{$#2#3$}}\kern-.5\wd0}
\makeatother

\begin{document}

\graphicspath{{figures/}}

\begin{center}

{\LARGE\bfseries
Thermodynamic Bethe Ansatz for the Toda chain from inverse scattering transform
}

\vspace{0.7cm}

{\large Alberto Brollo}

\vspace{0.3cm}

{\normalsize
Technical University of Munich, CIT, Department of Mathematics\\
Boltzmannstraße 3, 85748 Garching, Germany
}

\vspace{0.1cm}

{\normalsize
Technical University of Munich, Department of Physics\\
James-Franck-Straße 1, 85748 Garching, Germany
}

\end{center}

\vspace{0.5cm}

\begin{abstract}
The Thermodynamic Bethe Ansatz (TBA) procedure developed by Yang-Yang provides  the free energy functional to be minimized via a variational problem.  Their result not only gives the correct thermodynamics, but it also sets the framework for Generalized Hydrodynamics. Originally developed for  quantum systems, this procedure has been recently successfully applied to (integrable) classical particle systems. However, this has never been  done for classical field theories: no classical derivation has consistently  accounted for solitons, breathers, radiation (dispersive modes), and  their mutual interactions. We here derive the TBA equations for the Toda chain directly from its inverse scattering transform. This results in solitonic modes interacting with radiation, extending the soliton gas picture. We provide numerical verification of our equations against well established TBA equations in terms of particle modes. We expect the same procedure should apply to KdV, and pure radiative models as defocusing NLS and Sinh-Gordon, while we discuss current limitations in other models such as focusing NLS (and Ablowitz-Ladik) and Sine-Gordon. 
\end{abstract}

\tableofcontents

%==========================================================
%==========================================================
%==========================================================
\section{Introduction}
In recent years we have witnessed a large amount of scientific production in the field of integrable many-body systems~\cite{calabrese2016introduction,doyon2020lecture,bastianello2022introduction,Spohn24,doyon2025generalized}. This is mainly about one-dimensional Hamiltonian systems, and there are very good reasons for such interest. Integrable systems are, exceptionally, exactly solvable, and they allow to explore strongly interacting physics beyond approximations. This, combined with numerical accessibility of one-dimensional systems and experimental realizability~\cite{amerongen08,vogler13,schemmer2019generalized,malvania2021generalized} creates the perfect conditions for a fruitful physical program.
Although the early work of Yang--Yang goes back more than 50 years~\cite{Yang69,Takahashi99}, and Generalized Hydrodynamics~\cite{bertini2016transport,castro2016emergent} has by now a decade of history, the subject is far from being exhausted.

The Thermodynamic Bethe Ansatz (TBA) is the central framework for many-body integrable systems~\cite{Franchini_2017}. These are described by stable modes that scatter elastically, and the TBA connects the macroscopic density of these modes with the stationary states, the generalized Gibbs ensembles (GGEs). Generalized Hydrodynamics is naturally formulated in the same language.
The TBA was originally derived for quantum systems. However, its interpretation is simple and not specifically quantum, and this makes it reasonable that the same description applies to classical systems.
The difficulty is practical: in the quantum case the Bethe ansatz quantizes the spectrum, and this provides a way to count the modes and obtain their entropy, while for classical systems there is no quantization, and it is not obvious how the counting should be done.
For classical particle systems this has been achieved, first for the Toda fluid~\cite{Spohn19,Doyon19} and then for integrable particle systems with arbitrary scattering shift~\cite{doyon23}, and even for more exotic systems, as the box-ball system~\cite{kuniba2020generalized}. 

Different is for integrable nonlinear wave equations, or classical field theories, such as Korteweg de Vries, nonlinear Schr\"odinger, sine-Gordon, Landau-Lifshitz (as well as their lattice versions).
Since the late 1960s, the deterministic evolution of integrable nonlinear waves has been described via the inverse scattering transform (IST)~\cite{AblowitzSegur81,NMPZ84,Faddeev87}, obtaining exact profiles in terms of solitons and dispersive radiation. Much less is understood when the initial conditions are random~\cite{zakharov2009turbulence,El21}.
They feature as excitations solitons, breathers and dispersive radiation, and the obstacle to a TBA formulation is the counting of the modes of a wave field, in particular for delocalized modes as the dispersive radiation and extended solitons. It is not even clear whether radiation enters the thermodynamics at all, since in some semiclassical limits it is found to be replaced by extended solitons~\cite{koch22,Koch23}.
Existing approaches are indirect or partial. The Thermodynamic Bethe Ansatz has been obtained for some models through semiclassical limits of quantum ones~\cite{de2016equilibration,del2020exact,koch22,Koch23,bastianello24}, but not from first-principles.
Soliton gas theory~\cite{Zakharov71kinetic,el2003thermodynamic,el2005kinetic} works instead at the hydrodynamic level: it provides a kinetic equation for large numbers of solitons, with the same structure of Generalized Hydrodynamics~\cite{yoshimura18,bonnemain22}, but it excludes the dispersive modes, and whether it is complete is not settled.

Providing a TBA formulation for these systems is an old problem. The importance of the interplay between solitons and radiation was first noticed in the mid-seventies on the $\phi^4$ theory~\cite{Krumhansl75}, and a large number of studies followed in the eighties, mostly on sine-Gordon~\cite{Timonen86,TimonenBulloughPilling86,Bullough86,Claro1985}. They combined phenomenological soliton gases with semiclassical limits, and comparing with the exact free energy from transfer-integral methods. Agreement was found only in some regimes, and no definite conclusion was reached~\cite{Chung90}. The story is nicely summarized in Ref.~\cite{Koch23}.

\begin{figure}[tbp]
    \centering
    \includegraphics[width=\linewidth]{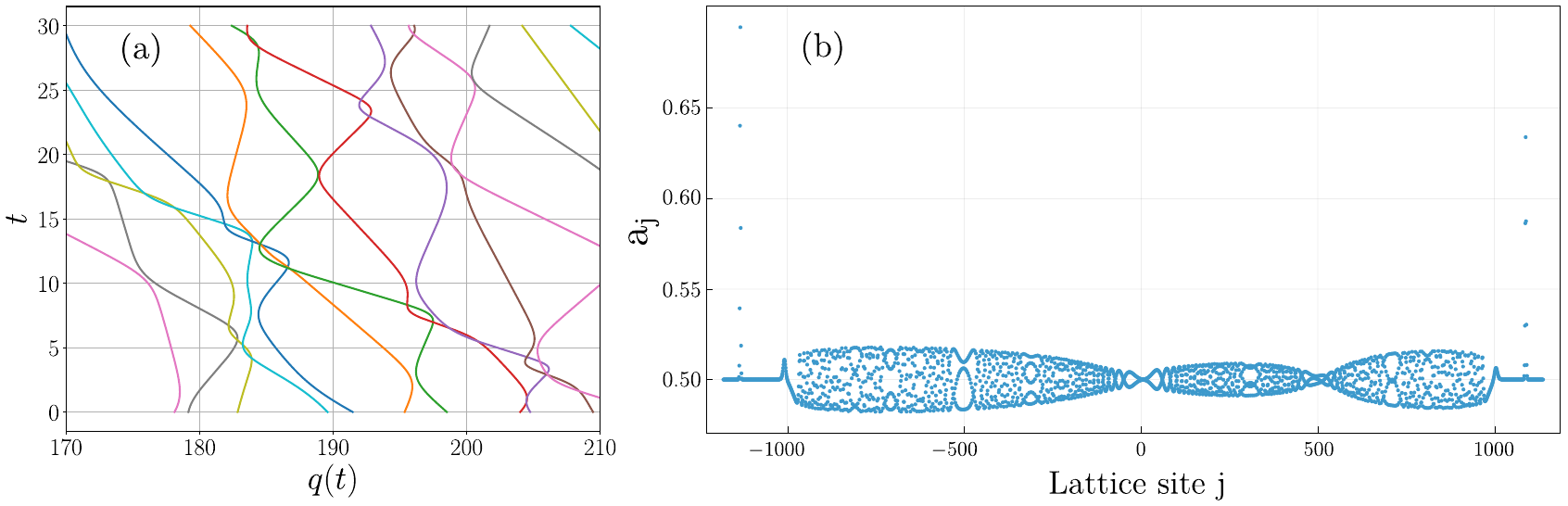}
    \caption{\textbf{Toda chain and Toda fluid.---} The same equations of motion~\eqref{eq:eom} admit a \emph{fluid} and a \emph{chain} picture, which differ in the interpretation of $q_j$. (a)~Fluid picture: $q_j$ is the position of the $j$-th particle, and the system is a gas of classical particles moving on the real line. (b)~Chain picture: $q_j$ is the displacement of the $j$-th particle from its lattice site, and excitations propagate on top of the chain vacuum, i.e.\ the infinite lattice at rest, $a_j\equiv 1/2$, $p_j\equiv 0$. The dynamics is typical of classical wave equations, with solitons and dispersive radiation. Shown is the Flaschka variable $a_j=\tfrac12e^{-r_j}$ at $t=1000$, obtained by solving~\eqref{eq:eomfl} with initial data $a_j(0)=\bigl|\tfrac12 - j e^{-j^2+j}\bigr|$ and $p_j(0)=-2j\,\mathrm{sech}\,j$. More detailed simulations of the Toda chain are reported in Ref.~\cite{Bilman2017}. The figure in panel~(a) is taken from Ref.~\cite{chahal2026}.}
    \label{fig:intro}
\end{figure}

A central model in this landscape is the Toda system~\cite{Toda81}
\be
H_\text{Toda} = \sum_j \frac{p_j^2}{2} + e^{-(q_{j+1}-q_j)}\, ,
\ee
with $q_j,p_j$ real canonically conjugate coordinates.
It admits two realizations that differ in the notion of space. In the \emph{Toda fluid}, space is the position $q_j$ of the particles, and the excitations are the particles themselves; in the \emph{Toda chain}, space is the label $j$, and the system behaves as a nonlinear wave on a lattice, with solitons and dispersive waves (radiation). See Fig.~\ref{fig:intro}.

From the fluid point of view the TBA formulation is well understood, and it consists of a single particle species with Maxwell-Boltzmann statistics, and we will refer to this as \emph{fluid TBA}. The first purely classical derivation is due to Spohn~\cite{Spohn19}, who connected the problem to the Dumitriu-Edelman $\beta$-ensembles of random matrix theory~\cite{dumitriu2002matrix}. Then Doyon~\cite{Doyon19} managed to count the entropy using the scattering structure of the theory, originally understood by Moser~\cite{Moser75}. Recently, a rigorous proof has also been obtained~\cite{grava2026large}. Historically, this was the first classical model whose correct TBA was derived from the semiclassical limit of its quantum counterpart~\cite{theod83,theod84,Takayama86}.

Since the TBA from the fluid side is very well understood, the Toda chain is an ideal setup to understand how to construct the TBA for nonlinear waves. In addition, it features a single species of solitons interacting with radiation, and no breathers, which makes it a minimal model for the soliton-radiation interaction.

In this paper, we give the first derivation of the TBA for the Toda chain in terms of solitons and radiation, based on the inverse scattering transform. We will refer to this as \emph{chain TBA}, to distinguish it from the \emph{fluid TBA} above.
Within this approach, the key object is the Lax operator, whose spectrum encodes the excitations: the discrete spectrum corresponds to solitons, the continuous spectrum connects to radiation.
The main new ideas we introduce are essentially two. The first one is a criterion to decide whether radiation is extensive or not. We show that the extensivity of the radiation is related to the localization of the absolutely continuous spectrum of the Lax operator, due to the randomness of the generalized Gibbs ensemble. In the case of Toda, the Lax operator takes the form of a random Jacobi operator, so this is really Anderson localization.
Then, we propose an ansatz for the counts of the entropy of the radiative modes, obtaining a Rayleigh-Jeans type of statistics. We further provide a numerical check of our derived \emph{chain TBA} equations against the well established \emph{fluid TBA} ones.

Although we focus on Toda, the approach is not specific to it, and in Sec.~\ref{sec:others} we discuss which parts extend to other models and where the limitations lie. We expect the same procedure to work, with minor modifications, for the Korteweg de Vries equation, whose direct scattering problem is Schr\"odinger. The method currently runs into difficulties for models whose direct scattering problem is the non-Hermitian Zakharov-Shabat problem, such as focusing NLS and sine-Gordon. In these models, solitons can be arbitrarily extended, and this changes drastically their entropy. Counting the phase-space volume occupied by extended solitons is, in our view, the central open problem in extending the present result. Nevertheless, a remarkable outcome is that the localization criterion applies also to these models, and shows that radiation is not extensive, in agreement with the absence of radiative modes in the TBA equations obtained from semiclassical limits~\cite{koch22,Koch23}.

\paragraph{Structure of the paper.}
Section~\ref{sec:toda} introduces the Toda chain, its solitons and radiation, the conserved charges and the scattering shifts. Section~\ref{sec:tba} recalls the generalities of TBA formalism. Section~\ref{sec:results} states the main results: the coupled soliton-radiation TBA equations, the localization criterion for the extensivity of radiation, and the Rayleigh-Jeans statistics of the radiative modes, together with the comparison with the fluid TBA. The derivation of the results occupies the following sections: Sec.~\ref{sec:ist} gives a self-contained account of the inverse scattering transform on the infinite line, Sec.~\ref{sec:scattering} derives the scattering shifts between solitons and radiation, and Sec.~\ref{sec:pf} derives the coupled TBA equations from the partition function expressed in action-angle variables. Section~\ref{sec:others} discusses the extension to other models, Korteweg-de Vries and the sinh- and sine-Gordon equations.

%==========================================================
%==========================================================
%==========================================================
\section{Toda Chain}\label{sec:toda}
The Toda chain describes the motion of a one-dimensional chain of particles coupled to their nearest neighbours through an exponential potential~\cite{Toda81}. The Hamiltonian takes the form
\be\label{eq:todaham}
H_\text{Toda} = \sum_j \frac{p_j^2}{2} + e^{-r_j}\, ,
\ee
with $q_j,p_j$ real canonically conjugate coordinates, and $r_j=q_{j+1}-q_j$ is called the \emph{stretch}. We use units such that the mass and the decay rate of the potential are equal to 1. The equations of motion are
\be\label{eq:eom}
\ddot{q}_j = e^{-(q_j - q_{j-1})} - e^{-(q_{j+1} - q_j)}~.
\ee
The Toda system admits two different pictures that differ for the notion of space~\cite{Doyon19}. In the \emph{chain} picture the physical space is made of the lattice sites $j$, and the $q_j$ are the displacements from the rest position. This is a simple model of a solid with nearest neighbour interaction, and this way was originally conceived by Toda~\cite{todaoriginal}, who already knew of KdV solitons and searched for an analogous feature in a discrete system.
Conversely, in the \emph{fluid} picture the physical space is the real line, and $q_j$ is the position of the $j$-th particle. This describes a gas of labeled classical particles that interact only with their neighbours in index space via an exponential repulsion.
A graphic representation of the two pictures is given in Fig.~\ref{fig:intro}. 
These two pictures come with different vacua: in the fluid picture a finite number of particles travel in an otherwise empty space (fluid vacuum); in the chain picture, excitations propagate on top of the infinite lattice at rest (chain vacuum). Any uniform stretch $r_j\equiv r_0$ is a stationary solution of~\eqref{eq:eom}, we fix $r_0=0$ by convention. Throughout this work we adopt the chain picture.

It is convenient to introduce the \emph{Flaschka variables} $a_j = \tfrac{1}{2}e^{-r_j/2}$, in terms of which the equations of motion are rewritten as
\be\label{eq:eomfl}
\dot{a}_j = \frac12 a_j(p_j - p_{j+1})~, \qquad
\dot{p}_j =  4(a_{j-1}^2 - a_{j}^2)~\, ,
\ee
and the chain vacuum is given by $a_j\equiv1/2$ and $p_j\equiv0$.

We now consider an infinite chain whose initial conditions decay rapidly to the vacuum: $a_j \to \tfrac{1}{2}$ and $p_j \to 0$ faster than any polynomial as $|j| \to \infty$. 
In this case, the system features solitonic solutions of the type
\be\label{eq:soliton}
a_j(t) = \frac{1}{2} \sqrt{\,1+\sinh^2\!\eta\;\operatorname{sech}^2\!\big(2|\eta| j - 2\sinh\eta\,t - \delta\big)\,}\,,
\ee
with \emph{rapidity} $\eta\in\mathbb{R}$ and $\delta$ set the initial position. This solution represent a non-dispersive solitary wave moving with velocity $v_s(\eta)=\,|\eta|^{-1}\sinh\eta$, $|v_s(\eta)|>1$. 
Looking at the example dynamics in Fig.\ \ref{fig:intro}, we see that the system features also dispersive solutions. For $|j/t|<1$ the solution fluctuates, although a pattern emerges. Fluctuations are enclosed by a slowly changing envelope that maintains its shape and spreads ballistically. This behavior is similar to that of linear wave equations, so the dispersive part of the solution is called \emph{radiation}.
At large times the radiation along the ray $j/t=\cos k$ is given by
\be\label{eq:asrad}
a_j(t) = \frac{1}{2} + \sqrt{\frac{\omega(k)A(k)}{8t}}\cos\left(jk + \frac12\omega(k)t + A(k)\log t - \delta^\pm(k)\right) + \mathcal{O}(t^{-1})\,,
\ee
where $\omega(k)=2\sin k$ is the frequency, $A(k)$ is a mode dependent amplitude fixed by the initial conditions, and $\delta^\pm(k)$ a mode dependent phase shift that distinguishes $t\to\pm\infty$ (more details in Sec.~\ref{sec:scattering}).

%inverse scattering
The system is completely integrable, in fact the equations of motion admit a Lax pair form. Specifically, they arise as the compatibility condition of the auxiliary linear problem
\be\label{eq:auxprob}
L(t)\psi(t) = \lambda\psi(t), \qquad
\frac{d}{dt}\psi(t) = B(t)\psi(t)\,,
\ee
where the Lax operators act as $(L f)_j = a_{j} f_{j+1} + \tfrac{1}{2}p_j f_j + a_{j-1} f_{j-1}$ and $(B f)_j = a_{j-1} f_{j-1} - a_{j} f_{j+1}$.
From now on, quantities without an explicit time dependence refer to initial-time values.
Introducing the spectral parameter $z \in \mathbb{C}$ via $\lambda = \tfrac{1}{2}(z + z^{-1})$, the first equation in \eqref{eq:auxprob} reads
\be\label{eq:laxprob}
a_{j-1}\psi_{j-1}(z)
+ \frac{1}{2}p_j\psi_j(z)
+ a_{j}\psi_{j+1}(z)
= \frac{z + z^{-1}}{2}\,\psi_j(z)\,.
\ee
This is a discrete Schr\"odinger-type problem with site dependent hopping and potential. In the asymptotic region $j\to \pm\infty$, where $a_j\to\tfrac12$ and $p_j\to0$, this reduces to the free tight-binding equation. This means that the spectrum is absolutely continuous for $|z|=1$, and discrete in the region $z_\alpha \in (-1,0) \cup (0,1)$. 

For $z=e^{ik}$ solutions are asymptotically plane waves, and a solution behaving as $\psi_j\sim e^{-ikj}$ for $j\to+\infty$, it goes as $\psi_j\sim \alpha(e^{ik})e^{ikj}+\beta_\mathrm{R}(e^{ik})e^{-ikj}$ for $j\to-\infty$. This defines the transmission coefficient $T(z)=1/\alpha(z)$ and the right reflection coefficient $R_\mathrm{R}(z)=\beta_\mathrm{R}(z)/\alpha(z)$.
The transmission coefficient can be meromorphically extended inside the unit disk $|z|<1$, with simple poles in correspondence of the point spectrum. It can be reconstructed via the Poisson-Jensen formula
\be\label{eq:T}
T(z) =
\left(\prod_{\alpha=1}^{M}
  \frac{|z_\alpha|(z - z_\alpha^{-1})}{z - z_\alpha}
\right)
\exp\!\left(
  \frac{1}{2\pi i}
  \int_{|z_k|=1} \log|T(z_k)|
  \frac{z_k + z}{z_k - z}\,\frac{dz_k}{z_k}
\right),
\qquad |z| < 1.
\ee
On the unit circle the integrand is singular, and the integral must be regularized via the Sochocki-Plemelj formula
\be\label{eq:sp}
\frac{1}{z_k - z\,e^{-0}}
= \operatorname{p.v.}\frac{1}{z_k - z}
  + \pi i\,\frac{\delta(z_k z^{-1}-1)}{z}\,,
\ee
where $\operatorname{p.v.}$ denotes the principal value. More details are in Sec.~\ref{sec:ist}.

The transmission coefficient is conserved under the dynamics and encodes the asymptotic behavior of the solution. At large times, a generic solution splits into two distinct spatial regions~\cite{Krueger09}: for $|j/t|<1$, dispersive radiation \eqref{eq:asrad} appears, with amplitude controlled by $A(k)=\frac{-\log|T(e^{ik})|^2}{\pi}$, while for $|j/t|>1$ one finds a collection of one-soliton solutions. Specifically, each pole $z_\alpha$ of $T(z)$ corresponds to a soliton with rapidity $\eta_\alpha = -\operatorname{sgn}(z_\alpha)\log|z_\alpha|$. See panel (b) of Fig.~\ref{fig:intro} for an example.

%scattering data
\paragraph{Action-Angle variables.} The complete information on the scattering problem is encoded in the \emph{scattering data}, which, besides $T(e^{ik})$ and $R_\mathrm{R}(e^{ik})$, include also the poles $z_\alpha$, and their right norming constant $c_{\mathrm{R},\alpha}$ (see Sec.~\ref{sec:ist}).
These are in one-to-one correspondence with the initial data.
In terms of the scattering data, we define a set of action-angle variables that diagonalize the Hamiltonian,
\be\label{eq:actionangle}
\begin{alignedat}{3}
  J(k) &= -\frac{\omega(k)}{2\pi}\log\lvert T(e^{ik})\rvert^{2},
    &\qquad \varphi(k,t) &= -\arg \frac{R_\mathrm{R}(e^{ik},t)}{T(e^{ik})},
    &\qquad k &\in (0,\pi), \\[2pt]
  \tilde p_\alpha &= z_\alpha + z_\alpha^{-1},
    &\qquad \tilde q_\alpha(t) &= \log\lvert c_{\mathrm{R},\alpha}(t)\rvert,
    &\qquad \alpha &= 1,\dots,M.
\end{alignedat}
\ee
and which satisfy the canonical Poisson brackets 
\be\label{eq:pb}
\{\varphi(k),\,J(k')\} = \delta(k-k')~, \qquad
\{\tilde q_\alpha,\,\tilde p_\beta\} = \delta_{\alpha\beta}~, 
\ee
while all other brackets vanish. In these variables the Hamiltonian
depends only on the actions, and the dynamics is linear
$\varphi(k,t)=\varphi(k,0)+\omega(k)t$ and $\tilde q_\alpha(t) = -(z_\alpha-z_\alpha^{-1}) t + \tilde q_\alpha(0)$.

%conserved charges
\paragraph{Conserved charges.} The transmission amplitude $T(z)$ is itself a conserved quantity, but it is non-local. Local charges are obtained by expanding $\log T(z)$ around $z=0$, precisely $Q_n=\tfrac{1}{n!}\p_z^n\log T(z)|_{z=0}$. Using the explicit form \eqref{eq:T}, the charges separates into a solitonic and a radiative component. In inverse scattering theory these are known as \emph{trace identities}~\cite{Toda81}
\be\label{eq:consch}
Q_n = \sum_{\alpha=1}^{M} \frac{z_\alpha^{-n} - z_\alpha^{n}}{n}
    + \int_0^\pi dk\; \frac{-\log|T(e^{ik})|^2}{\pi}\,(-\cos(nk)),
\ee
Notice that $\frac{-\log|T(e^{ik})|^2}{\pi}=A(k)$ is the radiation amplitude, while $q_n(k)=-\cos(nk)$ is the radiation charge density.
The first three charges are the total stretch, total momentum and the excitation energy above the vacuum 
\be\label{eq:RPE}
\begin{split}
  R & = \sum_{j=-\infty}^{+\infty} r_j = -2\log|T(0)|
    = \sum_{\alpha=1}^{M} -2|\eta_\alpha|
    + \int_0^\pi dk\;
      A(k), \\
  \mathcal{P} & = -Q_1
    = \sum_{\alpha=1}^{M} 2\sinh\eta
    + \int_0^\pi dk\;
      A(k)\,\cos k, \\
  E - E_0 & = Q_2
    = \sum_{\alpha=1}^{M} \sinh2|\eta_\alpha|
    + \int_0^\pi dk\;
      A(k)\,(-\cos 2k).
\end{split}
\ee
with $E_0$ the vacuum energy.

%scattering shifts
\paragraph{Scattering shifts.} A key property of integrability is the factorization of scattering. Toda already observed that solitons, despite the nonlinearity of their interaction, emerge from collisions unchanged except for a positional shift, and that this shift is a sum of two-body contributions. The rigorous long-time asymptotic obtained through the Riemann-Hilbert problem~\cite{Krueger09} extended this to the interaction of solitons with radiation. In Sec.~\ref{sec:scattering} we show that the same structure holds for the shift experienced by the radiation itself. Here we only state the result.

All the information on the scattering is encoded in the transmission coefficient~\eqref{eq:T}. We define the \emph{partial transmission coefficient}
\be\label{eq:pT}
\tilde T(z) =
\left(\prod_{\substack{\alpha=1 \\ z_\alpha \neq z}}^{M}
  \frac{|z_\alpha|(z - z_\alpha^{-1})}{z - z_\alpha}
\right)
\exp\!\left(
  \operatorname{p.v.}\int_0^\pi dk\;
  \frac{-\log|T(e^{ik})|^2}{\pi}\,
  \frac12\frac{z - z^{-1}}{(z + z^{-1}) - 2\cos k}
\right)~,
\ee
which is obtained from $T$ by removing the "self-contributions": the divergent factor when $z$ is a pole, or the delta contribution when $|z|=1$. Moreover, we rewrote the integral using that $T(e^{-ik})=\overline{ T(e^{ik})}$. 

Comparing the asymptotics at $t\to\pm\infty$, the soliton corresponding to the pole $z_\alpha$, of rapidity $\eta_\alpha$, receives the total shift $\Delta_\alpha = (2|\eta_\alpha|)^{-1}\log\tilde T(z_\alpha)^2$, while the radiation receives the mode-dependent shift $\Delta(k)=-\p_k\arg\tilde T(e^{ik})=i\,\p_k\log \tilde T(e^{ik})$, the last equality holding  since $|\tilde T(e^{ik})|=1$. Both shifts decompose into two-body contributions
\be\label{eq:scatt}
\begin{aligned}
&\Delta_\alpha = \sum_{\beta\neq\alpha}^M\Delta_{\rm ss}(\eta_\alpha,\eta_\beta) + \int_0^\pi dk\,A(k)\Delta_{\rm sr}(\eta_\alpha,k)\,,\\
&\Delta(k)=\sum_{\beta=1}^M\Delta_{\rm rs}(k,\eta_\beta) +\, \int_0^\pi dk'\,A(k')\Delta_{\rm rr}(k,k')
\end{aligned}
\ee
with two-body scattering shifts
\begin{equation}\label{eq:solshifts}
\begin{gathered}
\Delta_\mathrm{ss}(\eta,\eta')
= \frac{1}{2|\eta|}
\begin{cases}
\log\!\left(\dfrac{\sinh[(\eta+\eta')/2]}{\sinh[(\eta-\eta')/2]}\right)^{\!2}
& \eta\eta' > 0\, , \\[6mm]
\log\!\left(\dfrac{\cosh[(\eta-\eta')/2]}{\cosh[(\eta+\eta')/2]}\right)^{\!2}
& \eta\eta' < 0\, ,
\end{cases} \\[4mm]
\Delta_\mathrm{sr}(\eta,k)
= \frac{1}{2|\eta|}\,
\frac{-\sinh|\eta|}{\cosh\eta + \operatorname{sgn}(\eta)\cos k}\, ,
\end{gathered}
\end{equation}
\be\label{eq:radshifts}
\Delta_{\rm rr}(k,k')
    \;=\;\frac{d}{d k}\,\text{p.v.}\,\frac12\,\frac{-\sin k}{\cos k-\cos k'}\, ,
\qquad
\Delta_{\rm rs}(k,\eta)
\;=\;\frac{\sinh|\eta|}{\cosh\eta+\operatorname{sgn}(\eta)\cos k}\, .
\ee
The radiation-radiation shift $\Delta_{\rm rr}(k,k')$ has to be understood in the principal-value sense: the derivative acts on the regularized integral as a whole, which tames the non-integrable singularity at $k'=k$.

%==========================================================
%==========================================================
%==========================================================
\section{Thermodynamic Bethe Ansatz}
\label{sec:tba}
We now move on from the discussion of the Toda chain to present the Thermodynamic Bethe Ansatz (TBA), the general scheme describing many-body integrable systems at equilibrium. 
This was originally derived for quantum mechanical systems and it  is formulated in particle language, and not in wave (field theory) language~\cite{Franchini_2017}.
Its inputs are those introduced in the previous section: the stable excitations (in our case solitons and radiation), their mutual scattering shifts, and the conserved charges. However, a first principles derivation of TBA for inverse scattering solvable systems has never been carried out, and it is not clear how solitons and radiation should enter this formulation. In this paper, we will present this derivation for the Toda chain. We first review, in this section, how TBA generally works.

The starting point are the Generalized Gibbs ensembles (GGEs), the equilibrium states of integrable many body systems. While generic Hamiltonian systems conserve only mass, momentum and energy, integrable models possess an extensive number of local conserved charges $Q_n=\int dx\,q_n(x)$. All of them participate in the equilibrium state, extending the standard grand canonical measure from just energy and mass to all the charges
\be\label{eq:1pt}
\rho_\text{GGE} = \frac{1}{Z}\exp(-\sum_n \beta_nQ_n)~.
\ee
with $\beta_n$ the generalized temperatures, or Lagrange parameters, one for each conserved charge. 

The state of an integrable system is conveniently described in terms of stable modes that scatter elastically. A mode is specified by a label for the type $a$, which distinguishes qualitatively different excitations, and by a rapidity $\lambda_a\in\mathcal{D}_a$, which parametrizes the modes withing a given type. The set of all pairs $(a,\lambda_a)$ is the spectral space of the model, and sums over modes take the form $\sum_a\int d\lambda_a$, where the integration is over $\mathcal{D}_a$. What the spectral space is depends drastically on the model: in the Toda chain considered here, the two stable modes are the solitons and the radiation, whose rapidities live on the real line and on a Brillouin zone, respectively.
Each type carries a bare energy $e_a(\lambda_a)$, a bare momentum $p_a(\lambda_a)$, and also a "mass" $m_a$, in the sense of a non-dynamical charge: for spin chains this is the magnetization, for the Toda chain the stretch. More generally, each conserved charge $Q_n$ has a one-particle eigenvalue $h_{n,a}(\lambda_a)$, with $h_{n,a}\in\{m_a,p_a,e_a,\ldots\}$. A macrostate is described by the densities of states $\rho_a(\lambda_a)$, the number of modes of type $a$ per unit length with rapidity in $(\lambda_a,\lambda_a+d\lambda_a)$, and the charge densities are obtained by weighting the bare charges with them
\be
 q_n  = \sum_a\int d\lambda_a\,h_{n,a}(\lambda_a) \rho(\lambda_a)\,.
\ee
The goal of TBA is to connect GGEs to the corresponding densities of states: the former determines how much each conserved charge participates into the equilibrium states, the latter gives the local density for each type of mode in terms of the rapidity.

%interactions
The interactions are encoded in the two-body scattering shifts $\Delta_{ ab}(\lambda_a,\lambda_b')$, i.e. the displacement of the trajectory of mode $(a,\lambda_a)$ due to a collision with mode $(b,\lambda_b')$. For the Toda chain these are expressed in \eqref{eq:solshifts} and \eqref{eq:radshifts}. In the TBA, however, the shifts always appear multiplied by the derivative of the bare momentum, in the \emph{scattering kernel}
\be\label{eq:ker}
T_{ ab}(\lambda_a,\lambda_b') = \p_{\lambda_a}p_a(\lambda_a)\Delta_{ ab}(\lambda_a,\lambda_b')\,.
\ee
This structure originates in quantum systems, when the thermodynamic limit of the Bethe equations is taken. In classical mechanical systems, where $p=m\lambda$, this is usually also the case. However, in a generic classical system it may happen $\p_{\lambda_a}p_a(\lambda_a)$ not be the derivative of the physical momentum carried by the mode. For instance, for both Toda and KdV solitons it turns out to be the bare stretch $\p_{\lambda_a}p_a(\lambda_a)=\nu_a(\lambda_a)$ (see Ref.~\cite{bonnemain22} for a complete discussion). The choice is nevertheless not arbitrary, and it is fixed by the action-angle variables of the model. For radiative modes of classical field theories, instead, it works as for quantum excitations, due to an underlying classical analogue of the Bethe equations, as will be shown later.
It is possible for $\p_{\lambda_a}p_a(\lambda_a)$ to be negative, although always with a $\lambda_a$-independent sign. This is accounted for by the \emph{parity}
\be
\kappa_a = \operatorname{sgn}(\p_{\lambda_a}p_a)\,.
\ee

%dressing
The kernel defines the dressing operation. Given a rapidity resolved function $g_a(\lambda_a)$, its dressing $g_a\dr(\lambda_a)$ is the solution of the linear integral equation
\be
g_a\dr(\lambda_a) = g_a(\lambda_a) - \sum_{\rm b} \int\,d\lambda_b\,T_{\rm ab}(\lambda_a,\lambda_b)\kappa_b\vartheta_b(\lambda_b)g_b\dr(\lambda_b)\, ,
\ee
where $\vartheta_a(\lambda_a)$ is the filling function. We define the total density $\rho_a^{\bf t}=\kappa_a(\partial_{\lambda_a} p_a)\dr$. The filling function is the ratio $\vartheta_a=\rho_a/\rho_a^{\bf t}$. Combining the last two definitions, the densities of states and the total mode densities are related by
\be\label{eq:rhot}
\kappa_a\rho_a^{\bf t}(\lambda_a) = \p_{\lambda_a}p_a(\lambda_a) - \sum_b \int\,d\lambda_b\,T_{\rm ab}(\lambda_a,\lambda_b)\rho_b(\lambda_b)\, .
\ee
Hence, either $\{\rho_a\}$ or $\{\vartheta_a\}$ provide a complete description of the macrostate.

%free energy minimization
The equilibrium densities of states are obtained by minimizing the free energy density
\be\label{eq:tbafe}
f[\{\rho_a\}] = \sum_a \int~d\lambda_a\,w_a(\lambda_a)\rho_a(\lambda_a) - s[\{\rho_a\}]
\ee
where $w_a(\lambda_a)=\sum_n \beta_nh_{n,a}(\lambda_a)$ defines the GGE. For a thermal states at inverse temperature $\beta$ and chemical potential $\mu$, it is $w_a=\beta e_a - \mu m_a$. The functional $s[\{\rho_a\}]$ is the thermodynamic entropy density, which accounts for the number of microscopic configurations compatible with a given macrostate. It has the general form
\be\label{eq:tbaentropy}
s[\{\rho_a\}] = \sum_a \int~d\lambda_a\,\rho_a^{\bf t}(\lambda_a) s_a(\vartheta_a(\lambda_a))
\ee
where $s_a(\vartheta)$ distinguishes the different statistical weights of the different modes~\cite{doyon2020lecture}. 
In classical systems one typically encounters the Maxwell-Boltzmann functional form $s_\text{MB}(\vartheta) = \vartheta(1-\log\vartheta)$ for classical particles and Rayleigh-Jeans form $s_\text{RJ}(\vartheta)=\log\vartheta+1$ for radiation. The Maxwell-Boltzmann form would be expected to apply to solitons as well, but recent results from semiclassical limits show this to be false when accounting for very extended solitons (see Sec.~\ref{sec:others}). 

The functional minimization of the free energy density \eqref{eq:tbafe} gives the nonlinear integral equations known as TBA equations
\be
s_a'(\vartheta_a(\lambda_a)) = w_a(\lambda_a) - \sum_b \int\,d\lambda_b\, T_{\rm ab}(\lambda_a,\lambda_b)\kappa_b\Big[\vartheta_b s_b'(\vartheta_b) - s_b(\vartheta_b)\Big](\lambda_b)
\ee
with $s_a'(\vartheta)=ds_a/d\vartheta$. Typically, one introduces the \emph{pseudoenergy} $\varepsilon_a=s_a'(\vartheta_a)$, and the TBA equations become integral equation for $\varepsilon_a$. Their numerical solution is a standard task, and once the pseudoenergies are known, the filling functions and the densities of states follow.
The average charges are obtained by \eqref{eq:1pt}, and by standard thermodynamical relations, the connected correlators of the charge densities $\langle q_mq_n\rangle_\text{c}=\langle q_mq_n\rangle-\langle q_m\rangle\langle q_n\rangle$ are obtained deriving the expectation values with respect to the generalized temperatures
\be\label{eq:corr}
\langle q_mq_n\rangle_\text{c} = -\frac{\p}{\p\beta_m}\langle q_n\rangle = \sum_a \int~d\lambda_a\,\rho_a(\lambda_a)\chi_a(\lambda_a)h_{n,a}\dr(\lambda_a)h_{m,a}\dr(\lambda_a)\,, \qquad
\chi_a = -\frac{1}{\vartheta_as''(\vartheta_a)}\,,
\ee
where $\langle\ldots\rangle$ is the GGE average and $\chi_a$ a statistical factor. Explicitly, for Maxwell-Boltzmann statistics $\chi_a=1$ and for Rayleigh-Jeans $\chi_a=\vartheta_a$. 

%ghd
\paragraph{Generalized Hydrodynamics (GHD).} This is the theory for nonequilibrium integrable systems at large space-time scales. At Euler scale, the system is locally in equilibrium, so GHD simply extends TBA to nonequilibrium smoothly inhomogeneous states. The densities of states become space-time dependent $\rho_a(\lambda_a;x,t)$, and the dynamics follows from the conservation laws $\p_t\langle q_n\rangle+\p_x\langle j_n\rangle=0$. The average currents in a GGE are~\cite{bertini2016transport,castro2016emergent,yoshimura2020collision,borsi2021current}
\be
\langle j_n\rangle = \sum_a \int~d\lambda_a\, h_{n,a}(\lambda_a)v_a^\text{eff}(\lambda_a)\rho_a(\lambda_a)\,,\qquad
v_a^\text{eff} = \frac{(\p_{\lambda_a}e_a)\dr}{(\p_{\lambda_a}p_a)\dr}\,,
\ee
where $v_a^{\rm eff}$ is the \emph{effective velocity}, the velocity at which a mode propagates through the medium. As we previously discussed for the momentum derivative, $\p_{\lambda_a}e_a(\lambda_a)$ may not be the derivative of the physical energy. Once the momentum derivative is fixed, it holds $\p_{\lambda_a}e_a(\lambda_a)=v_a(\lambda_a)(\p_{\lambda_a}p_a(\lambda_a))$, with $v_a(\lambda_a)$ the bare velocity of the mode, that is an unambiguous physical quantity.

Manipulating the dressing equations, the effective velocity can equivalently be written as the solution of
\be\label{eq:veff}
v_a^{\rm eff}(\lambda_a)=v_a(\lambda_a)+\sum_b\int d\lambda_b\,\Delta_{ab}(\lambda_a,\lambda_b)\,\rho_b(\lambda_b)\big[v_a^{\rm eff}(\lambda_a)-v_b^{\rm eff}(\lambda_b)\big]\,,
\ee
This is the form originally derived for soliton gases~\cite{El21}, also known as \emph{collision rate ansatz}. 
Here, though, we will show this to be more general, applying the same way not only to solitons, but also to radiation.

Since the charges form a complete set, the conservation laws imply the GHD equation
\be\label{eq:GHD}
\p_t\rho_a(\lambda_a;x,t)+\p_x\Big[v_a^{\rm eff}(\lambda_a;x,t)\rho_a(\lambda_a;x,t)\Big]=0\,,
\ee
a nonlinear continuity equation for each mode, the nonlinearity entering only through $v^{\rm eff}_a$. 
The above GHD equation may be used also to compute the dynamical correlations of the charges in the linear response regime. These take the form~\cite{doyonspohn,doyoncorr}
\be\label{eq:eulercorr}
\langle q_n(x,t)q_m(0,0)\rangle_c=\sum_a\int d\lambda_a\,\rho_a(\lambda_a)\,\chi_a(\lambda_a)\,h_{n,a}\dr(\lambda_a)\,h_{m,a}\dr(\lambda_a)\,\delta\big(x-v_a^{\rm eff}(\lambda_a)\,t\big)\,.
\ee
The initial time correlations \eqref{eq:corr} are transported ballistically by the modes along rays corresponding to their effective velocity.

%==========================================================
%==========================================================
%==========================================================
\section{Main Results}\label{sec:results}
We now return to the Toda chain. In Sec.~\ref{sec:sr} we present our result for the free energy functional in terms of soliton and radiation densities. We derive it from the partition function by means of the inverse scattering transform; details are reported in Sec.~\ref{sec:pf}.
Once the free energy functional is obtained, the TBA scheme described in the previous section can be applied directly. The effective velocities follow, and the Euler-scale Generalized Hydrodynamics equations couple solitons and radiation, thereby extending the soliton gas picture. We refer to this formulation as the \emph{chain TBA}.

As discussed in Sec.~\ref{sec:toda}, the Toda system admits two pictures. In Sec.~\ref{sec:qp} we then present the \emph{fluid TBA}, formulated in terms of \emph{quasiparticles}, the scattering states in the fluid picture. This formulation is well established in the literature and has been extensively tested against molecular dynamics simulations~\cite{bulchandani19,cao2019gge,mazzuca23,chahal2026}. 
The chain and fluid pictures rely on different notions of space: densities and currents are defined per lattice site in the former and per unit length in the latter. The two are related through the stretch, and computing chain observables from the fluid TBA is routine~\cite{Doyon19}. 
We go further and show how the equilibrium states of the two formulations are connected, namely how a given quasiparticle density maps to the corresponding soliton and radiation densities. Finally, we show that the two formulations of Generalized Hydrodynamics agree at the level of the Euler-scale two-point correlators of the conserved charges. Since the fluid TBA is derived independently and is well established, this agreement provides a nontrivial check of the chain TBA.

\subsection{Chain TBA: coupled soliton-radiation system}
\label{sec:sr}
%partition function
In Sec.~\ref{sec:toda} we discussed the Toda chain with deterministic, rapidly decaying initial data. We now turn to random initial conditions. 
Consider the infinite Toda chain prepared in its vacuum, $a_j = 1/2$, $p_j = 0$, except for a region of $N$ consecutive sites, which is excited according to a generalized Gibbs ensemble with weight $W$. For simplicity, we express the GGE weight as a linear combination of the local conserved charges $W = \sum_n \beta_n Q_n$, although it can be of more general form. %check
Since the total stretch $R=\sum_{j=1}^{N}r_j$ is also conserved by the dynamics, we introduce its thermodynamic dual, the pressure $P$, where $P>0$ is required to have a finite value of the stretch. The finite volume partition function reads
\be\label{eq:ZL}
Z_N(\boldsymbol{\beta},P) = \int \prod_{j=1}^N dp_jdq_je^{-PR}e^{-W}
\ee
where $\boldsymbol{\beta}$ denotes the vector of Lagrange parameters $\beta_n$. Note that no factor of $1/N!$ appears since lattice sites are distinguishable. 

A convenient choice is to change variables to the initial time action-angle ones \eqref{eq:actionangle}. This introduces a trivial Jacobian, and since the charges separate exactly into solitonic and radiative contribution (cfr. \eqref{eq:consch}), so does also the GGE, $W = \sum_{\alpha} w_\alpha
+ \int_0^\pi dk\,A(k)\,w(k)$, with
\be\label{eq:gge}
w_\alpha=\sum_n\beta_n\,\frac{z_\alpha^{-n} - z_\alpha^{n}}{n}\,,
\qquad
w(k)=\sum_n\beta_n(-\cos(nk))\,,
\ee
and $A(k)$ the radiation amplitude dependent only from the radiation action variable $J(k)$.
Thus, the GGE depends only on the action variables, leading to a partial factorization of the partition function
\be\label{eq:ZList}
\begin{split}
&Z_N(\boldsymbol{\beta},P) = \sum_{M}\frac{1}{M!} \int \prod_{\alpha=1}^M d\tilde p_\alpha\exp\left(-\sum_{\alpha=1}^M w_\alpha + P\,\nu_\alpha\right)~\times \\
&\int\mathcal{D} J(k)\exp\left(-\int_0^\pi dk \,A(k)(w(k) + P)\right)\int \prod_{\alpha=1}^M d\tilde q_\alpha\int\mathcal{D}\varphi(k)
\end{split}
\ee
where $\nu_\alpha$ is the stretch carried by each soliton, while $\int_0^\pi dk\,A(k)$ that carried by the radiation, cfr. \eqref{eq:RPE}. The sum over the number of solitons appears since the phase space of the action-angle variables is made of disjoint sectors with different numbers of solitons.

In the thermodynamic limit $N \to \infty$, the mean soliton number grows extensively with $N$, consistent with a finite soliton density set by the GGE. The extensivity of the radiative sector is less immediate and requires separate justification.
When initial data are randomly sampled from a GGE, the bulk of the Lax operator becomes a random Jacobi operator. In one dimension, a stationary and ergodic disorder produces a positive Lyapunov exponent $\gamma(\lambda)$ for almost every $\lambda$, almost surely~\cite{TeschlToda}. The Lyapunov exponent sets an inverse localization length for the solutions of the eigenvalue equation $L\psi=\lambda\psi$. Where $\gamma(\lambda)>0$ the solution decays exponentially over a scale $1/\gamma(\lambda)$, if $\lambda$ belongs to the point spectrum. Only where $\gamma(\lambda)=0$ solutions can remain extended and oscillatory, contributing instead to the absolute continuous spectrum. A stationary and ergodic disorder produces localized solutions for almost every $\lambda$. This is the mechanism of Anderson localization.
In our setup, this means that the incoming and outgoing plane waves $e^{ikj}$ are connected by an exponentially suppressed transmission coefficient $|T(e^{ik})| \sim \exp\!\bigl(-N\,\gamma_N(k)\bigr)$, with $\gamma_N(k)$ the finite size Lyapunov exponent~\cite{kostrykin99} (see Sec.~\ref{sec:ist}). This is a fluctuating quantity, self-averaging $\gamma_N\to\gamma$ almost surely as $N\to\infty$, with Gaussian fluctuations of size $O(\sqrt{N})$.

This exponential suppression is precisely what makes the action variable extensive. At large but finite $N$ we have for the typical $J(k)$
\be\label{eq:extrad}
  J(k)
  = -\frac{\omega(k)}{2\pi}\log\bigl|T(e^{ik})\bigr|^2
  = N\frac{\omega(k)}{\pi}\,\gamma(k)\, + O(\sqrt{N}).
\ee
This argument for the extensivity of the radiation is, to the best of our knowledge, a new one and one of the central contribution of this work. In Sec~\ref{sec:others} we show that it successfully applies also to other models.

%free energy
The partition function \eqref{eq:ZList} is expressed in a factorized form, so the integration over the angle variables may appear trivial. However, the phase-space volume available to each soliton is renormalized by the interactions and depends on the action variables. At large $N$, this integration can be performed using a mean field argument, analogous to the one of Ref.~\cite{Doyon19}. For the radiation we instead discretize the measure $\mathcal DJ(k)\,\mathcal D\varphi(k)$ on the roots of a classical analogue of the Bethe equations, obtained from the periodic problem, and we assume the measure to be flat in the discretized variables. The interaction then enters through the density of the Bethe roots. We discuss the details in Sec.~\ref{sec:pf}.
Finally, in the thermodynamic limit $N\to\infty$, we introduce the coarse-grained soliton density, $\rho_\mathrm{s}(\eta)$, as a function of the soliton rapidity $\eta$, and the radiation density, $\rho_\mathrm{r}(k)$. 
The coarse-graining procedure intrinsically gives rise to an entropy: for particles and solitons this is known to be the Shannon entropy, whereas the corresponding expression for the radiation is a new result of this paper.

The total densities \eqref{eq:rhot} take the form
\begin{align}\label{eq:spacedensity}
&\rho_\mathrm{s}^\mathbf{t}(\eta) = 2|\eta|
- \int\,d\eta'\,T_\mathrm{ss}(\eta,\eta')\rho_\mathrm{s}(\eta')
- \int_0^\pi dk'\,T_\mathrm{sr}(\eta,k')\, \rho_\mathrm{r}(k')\, , \\
&\rho_\mathrm{r}^\mathbf{t}(k) = \frac{1}{\pi}
+ \int\,d\eta'\,T_\mathrm{rs}(k,\eta')\rho_\mathrm{s}(\eta')
+ \int_0^\pi dk'\,T_\text{rr}(k,k')\, \rho_\mathrm{r}(k') \, ,
\end{align}
meaning that momentum derivative functions are $\p_\eta p_\mathrm{s}(\eta)=2|\eta|$ and $\p_k p_\mathrm{r}(k)=-1/\pi$, and the scattering kernels are defined as in \eqref{eq:ker} starting from the scattering shifts \eqref{eq:solshifts} and \eqref{eq:radshifts} 
\be\label{eq:kernels}
\begin{split}
&T_\mathrm{ss}(\eta,\eta')=2|\eta|\Delta_\mathrm{ss}(\eta,\eta')\,,\qquad
T_\mathrm{sr}(\eta,k)=2|\eta|\Delta_\mathrm{sr}(\eta,k)\,, \\
&T_\mathrm{rs}(k,\eta)=-\frac{1}{\pi}\Delta_\mathrm{rs}(k,\eta)\,,\qquad
T_\mathrm{rr}(k,k')=-\frac{1}{\pi}\Delta_\mathrm{rr}(k,k')\,.
\end{split}
\ee
Then the filling functions are defined, $\vartheta_\mathrm{s}=\rho_\mathrm{s}/\rho_\mathrm{s}^\mathbf{t}$ and $\vartheta_\mathrm{r}=\rho_\mathrm{r}/\rho_\mathrm{r}^\mathbf{t}$.

After integrating out the angle variables and course-graining, the GGE measure takes the large deviation form $\mathbb{P}[\rho_\mathrm{s},\rho_\mathrm{r}] \asymp e^{-N(\mathcal{F}_S[\rho_\mathrm{s},\rho_\mathrm{r}]+\mathcal{F}_R[\rho_\mathrm{s},\rho_\mathrm{r}])}$, with free energies
\be\label{eq:freeE}
\begin{split}
&\mathcal{F}_S[\rho_\mathrm{s},\rho_\mathrm{r}] = \int d\eta\,\rho_\mathrm{s}(\eta)
\Big (w(\eta) - P(2|\eta|)
- \log|2\sinh\eta| + \log\vartheta_\mathrm{s}(\eta) - 1\Big), \\
&\mathcal{F}_R[\rho_\mathrm{s},\rho_\mathrm{r}] = \int_0^\pi dk\,\rho_\mathrm{r}^\mathbf{t}(k)
\Big[\big(w(k)+P\big)\vartheta_\mathrm{r}(k) - \log(\pi\omega(k)\vartheta_\mathrm{r}(k)) - 1\Big].
\end{split}
\ee
Two unconventional terms appear, $\log|2\sinh\eta|$ and $\log(\pi\omega(k))$. The former is due to a Jacobian resulting from reparametrization of the rapidity,, while the latter is due to a redefinition of the radiation density (see Sec.~\ref{sec:pf}). A part for that, the two free energies have the standard Maxwell-Boltzmann form, for solitons, and Rayleigh-Jeans form for radiation, as presented below \eqref{eq:tbaentropy}. The entropic terms arise as the combination of the renormalized phase space volume due to interactions, and the course-graining procedure.

The equilibrium configuration is provided by the joint minimization of the free energies, that produces the coupled TBA equations
\be\label{eq:TBA}
\boxed{
\begin{aligned}
\varepsilon_\mathrm{s}(\eta) &= w(\eta) - 2P|\eta| - \log|2\sinh\eta|
+ \int d\eta'\,T_\mathrm{ss}(\eta,\eta')\,e^{-\varepsilon_\mathrm{s}(\eta')}
+ \int_0^\pi dk\,T_\mathrm{rs}(k,\eta)\,\log\frac{\varepsilon_\mathrm{r}(k)}{\pi\omega(k)}\,,\\
\varepsilon_\mathrm{r}(k) &= w(k) + P
+ \int d\eta\,T_\mathrm{sr}(\eta,k)\,e^{-\varepsilon_\mathrm{s}(\eta)}
+ \int_0^\pi dk'\,T_\mathrm{rr}(k',k)\,\log\frac{\varepsilon_\mathrm{r}(k')}{\pi\omega(k')}\,,
\end{aligned}
}
\ee
with the pseudoenergies defined as $\vartheta_\mathrm{s}=e^{-\varepsilon_\mathrm{s}}$ and $\vartheta_\mathrm{r}=1/\varepsilon_\mathrm{r}$. 

We can then compute also the effective velocities, as in \eqref{eq:veff}
\be
\begin{split}
v_\mathrm{s}^{\rm eff}(\eta)&=v_\mathrm{s}(\eta)
+\int d\eta'\,\Delta_\mathrm{ss}(\eta,\eta')\,\rho_\mathrm{s}(\eta')\big[v_\mathrm{s}^{\rm eff}(\eta)-v_\mathrm{s}^{\rm eff}(\eta')\big]\\
&\quad+\int_0^\pi dk'\,\Delta_\mathrm{sr}(\eta,k')\,\rho_\mathrm{r}(k')\big[v_\mathrm{s}^{\rm eff}(\eta)-v_\mathrm{r}^{\rm eff}(k')\big]~,\\
v_\mathrm{r}^{\rm eff}(k)&=v_\mathrm{r}(k)
+\int d\eta'\,\Delta_\mathrm{rs}(k,\eta')\,\rho_\mathrm{s}(\eta')\big[v_\mathrm{r}^{\rm eff}(k)-v_\mathrm{s}^{\rm eff}(\eta')\big]\\
&\quad+\int_0^\pi dk'\,\Delta_\mathrm{rr}(k,k')\,\rho_\mathrm{r}(k')\big[v_\mathrm{r}^{\rm eff}(k)-v_\mathrm{r}^{\rm eff}(k')\big]~,
\end{split}
\ee
with $v_\mathrm{s}(\eta)=\sinh\eta/|\eta|$ the soliton bare velocity and $v_\mathrm{r}(k)=\omega'(k)/2=-\cos k$ the radiation bare velocity, obtained by taking the stationary phase in \eqref{eq:asrad}.
The GHD equations are then
\be
\p_t\rho_\mathrm{s}(x,t;\eta) + \p_x\big(v_\mathrm{s}^\mathrm{eff}\rho_\mathrm{s}\big)(x,t;\eta) = 0,\qquad
\p_t\rho_\mathrm{r}(x,t;k) + \p_x\big(v_\mathrm{r}^\mathrm{eff}\rho_\mathrm{r}\big)(x,t;k) = 0,
\ee
with the effective velocities evaluated on the local state, and $x$ the lattice coordinate $j$, treated as a continuous variable at the hydrodynamic scale.

In the case $\rho_\mathrm{r}=0$, one retrieves the soliton gas equations. This picture here is extended to a genuine two component hydrodynamics of solitons and radiation. Interestingly, the latter behaves on the same footing as solitons in the hydrodynamic description: the dispersive behavior does not seem to have a role, and only the factorization of the scattering matters, similarly to what happens in quantum models where quasiparticles are dispersive wavepackets of Bethe eigenstates~\cite{Vlijm_2015}.

\subsection{Comparison with the fluid TBA}\label{sec:qp}
The Toda system admits another TBA description, formulated in the fluid picture~\cite{Doyon19,Spohn19}, see~\cite{Spohn24} for a complete review. There, the stable excitations are \emph{quasiparticles}, labelled by the eigenvalues of the Lax matrix and interacting through a single two-body scattering shift. Here we recall only its essential elements, and then discuss the correspondence with the chain TBA of Sec.~\ref{sec:sr}. 

\paragraph{Toda fluid and quasiparticles.} We consider $N$ particles with open boundary conditions $a_0=a_N=0$. As the rapid decay to the vacuum in the chain picture, this choice guarantees that the dynamics has well-defined scattering states. The Lax matrix $L_N$ is the restriction of the Lax operator in \eqref{eq:auxprob} to the sites $j=1,\ldots,N$, i.e. a real symmetric Jacobi matrix with diagonal entries $p_j/2$ and off-diagonal entries $a_j$. The asymmetric partner matrix $B_N$ is obtained in the same way, and the equations of motion \eqref{eq:eom} take the Lax form $\dot{L}_N=[B_N,L_N]$. Hence $L_N(t)$ is isospectral and its eigenvalues $\lambda_1,\ldots,\lambda_N$ are conserved. They define the fluid charges
\be
Q^{\rm f}_n = 2\operatorname{Tr}L_N^n = 2\sum_{\alpha=1}^N \lambda_\alpha^n\, ,
\ee
where the superscript distinguishes them from the chain charges \eqref{eq:consch}. In the latter case, the Lax matrix is infinite dimensional, but trace class, and it turns out the chain conserved charges are $Q_n=2(\operatorname{Tr}(L)^n-\operatorname{Tr}(L_0)^n)$~\cite{Toda81}, so they differ from the fluid charges by removing the vacuum contribution.

The eigenvectors evolve as $\dot{\psi}_\alpha=B_N\psi_\alpha$, cfr. \eqref{eq:auxprob}, and allow one to define the positions~\cite{bulchandani19,lepri2025,aggarwal2026,chahal2026}
\be\label{eq:qppos}
Q_\alpha(t) = \sum_{j=1}^N q_j(t)|\psi_\alpha(j,t)|^2\,,
\qquad \dot Q_\alpha(t) = \langle\psi_\alpha(t)|\mathbb{V}_N(t)|\psi_\alpha(t)\rangle\,, 
\ee
with $\alpha=1,\ldots,N$, $\psi_\alpha(j,t)$ the $j$-th component of the normalized eigenvector at time $t$, and $\mathbb{V}$ the symmetric tridiagonal velocity matrix with entries $\mathbb{V}_{j,j} = p_j$,  $\mathbb{V}_{j,j+1} = \mathbb{V}_{j+1,j} =-2a_j\log 2a_j$. Since the interaction is repulsive, as $t\to\pm\infty$ the particles separate and $a_j(t)\to0$. Then both $L_N(t)$ and $\mathbb{V}(t)$ become diagonal, and $\dot Q_\alpha\to v_\alpha=2\lambda_\alpha$. Hence $Q_\alpha(t)$ is the trajectory of an object with given asymptotic velocity $v_\alpha$. This are the \emph{quasiparticles} of the Toda fluid, as originally discovered by Moser~\cite{Moser75}. Their scattering is elastic, and a collision with a quasiparticle of velocity $v'$ displaces its trajectory by
\be\label{eq:qpshift}
\Delta_\mathrm{qp}( v, v')=-2\log| v- v'|\,,
\ee
in the sign convention of \eqref{eq:scatt}, i.e. the displacement of the fastest.

The Lax matrix diagonalizes at large times, so the asymptotic eigenvectors are supported on a single component. From \eqref{eq:qppos}, this means that particles and quasiparticles asymptotically coincides. Hence, the quasiparticles rapidities $v_\alpha$ are given by $v_\alpha=\lim_{t\to+\infty}p_j(t)$. We also define the corresponding \emph{impact parameters} $q_\alpha=\lim_{t\to+\infty}q_j(t)-t\,p_j(t)$. Due to Liouville's theorem, the \emph{scattering coordinates} $(v_\alpha,q_\alpha)$ are action-angle variables for the open Toda fluid.

\paragraph{Fluid TBA.} In the fluid picture the length of the system is the total stretch $R=\sum_{j=1}^{N}r_j$, while $N$ is the number of particles. The natural ensemble fixes $R$ and lets $N$ fluctuate, with chemical potential $\mu$~\cite{Doyon19}
\be
Z^{\rm f}_R(\boldsymbol{\beta},\mu) = \sum_N e^{\mu N} \int \prod_{j=1}^N dp_jdq_j\delta\Big(\sum_{j=1}^{N}r_j - R\Big)e^{-W^{\rm f}}\,.
\ee
This is dual to the chain ensemble \eqref{eq:ZL}, which fixes the number of sites and weights the total stretch through the pressure $P$. The function $W^{\rm f}=\sum_n\beta_nQ_n^{\rm f}$ defines the GGE, expressed as linear combination of the fluid local conserved charges. A given set of Lagrange parameters $\boldsymbol{\beta}$ defines the same state in the chain and fluid ensemble, since the charges differ only for a constant shift.

In scattering coordinates the GGE depends only on the actions
\be
Z^{\rm f}_R(\boldsymbol{\beta},\mu) = \sum_N \frac{e^{\mu N}}{N!} \int \prod_{\alpha=1}^N dv_\alpha e^{-\sum_\alpha w^{\rm f}(v_\alpha)} \int \prod_{\alpha=1}^N dq_\alpha \delta\Big(\sum_{j=1}^{N}r_j - R\Big)\,,
\ee
with $w^{\rm f}(v_\alpha)=\sum_n \beta_n2^{1-n}v_\alpha^n$. The factor $1/N!$ now accounts for the arbitrary association between particles and quasiparticles. The integration over the angle variables is carried out at the mean field level as for the chain: the phase space volume available to each quasiparticle is renormalized by its scattering shifts with all the others~\cite{Doyon19} (see. Sec~\ref{sec:cpTBA}). Introducing the coarse-grained quasiparticle density $\rho_\mathrm{qp}(v)$ per unit length, one obtains $Z_R^{\rm f} \asymp \int\mathcal{D}\rho_\mathrm{qp}e^{-R\mathcal{F}_\mathrm{qp}[\rho_\mathrm{qp}]}$, where the normalization of $\rho_\mathrm{qp}$ is unconstrained since $N$ fluctuates, and
\be
\mathcal{F}_{\rm qp}[\rho_{\rm qp}] = \int dv \rho_{\rm qp}(v)\Big (w^{\rm f}(v) - \mu + \log\vartheta_\mathrm{qp}(v) - 1\Big)\,,
\ee
with
\be\label{eq:qprhot}
\rho_\mathrm{qp}^{\bf t}(v) = 1 - \int dv'T_\mathrm{qp}(v,v')\rho_\mathrm{qp}(v')\,,\qquad \vartheta_\mathrm{qp}(v)=\rho_\mathrm{qp}(v)/\rho_\mathrm{qp}^{\bf t}(v)\,.
\ee
This is the general strcuture of Sec.~\ref{sec:tba} with a single mode of Maxwell-Boltzmann statistics, $\p_v p_\mathrm{qp}=1$, and kernel $T_\mathrm{qp}=\Delta_\mathrm{qp}$, cfr. \eqref{eq:ker} and \eqref{eq:rhot}. The minimization yields the TBA equation
\be
\varepsilon_\mathrm{qp}(v) = w^{\rm f}(v) - \mu - 2\int dv'\log|v-v'|e^{-\varepsilon_{\rm qp}(v')}\,,
\qquad \vartheta_{\rm qp}=e^{-\varepsilon_{\rm qp}}\,.
\ee
The filling function is normalized to the pressure of the corresponding chain ensemble $\int dv\,\vartheta_\mathrm{qp}(v)=P$, which fixes $\mu$ for a given $P$. The mean stretch then follows from $\nu^{-1}=\int dv\,\rho_\mathrm{qp}(v)$, with $\nu=\langle r_0\rangle$.
The effective velocity follows from \eqref{eq:veff} with the shift \eqref{eq:qpshift},
\be
v^\text{eff}_{\rm qp}(v) = v - 2\int dv'\log|v-v'|\rho_\mathrm{qp}(v')\big(v^\text{eff}_{\rm qp}(v)-v^\text{eff}_{\rm qp}(v')\big)\,,
\ee
and the GHD equation reads $\p_t\rho_\mathrm{qp}+\p_x(v^\text{eff}_\mathrm{qp}\rho_\mathrm{qp})=0$, with $x$ the position on the real line.

\begin{figure}[tbp]
    \centering
    \includegraphics[width=\linewidth]{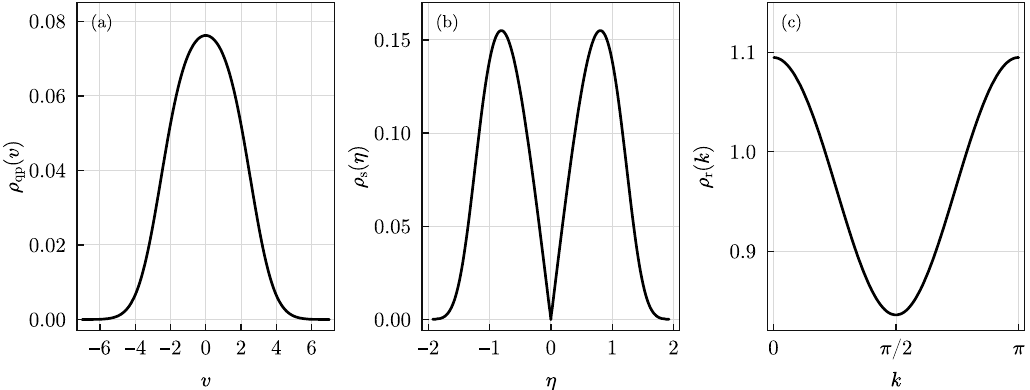}
    \caption{\textbf{Identification of the densities.}--- Densities of states of a thermal equilibrium state at $\beta=0.5$, $P=0.32$. (a) The state in terms of the quasiparticle density $\rho_\mathrm{qp}(v)$ of the fluid TBA. (b),(c) The equivalent description in terms of the soliton density $\rho_\mathrm{s}(\eta)$ and of the radiation density $\rho_\mathrm{r}(k)$ of the chain TBA, via \eqref{eq:soldens} and \eqref{eq:raddens}. The quasiparticles with $|v|>2$ map onto the solitons, while the radiation is connected to the Lyapunov exponent through the Thouless formula \eqref{eq:thouless}.}
    \label{fig:dos}
\end{figure}

\paragraph{Identification of the densities.} A macroscopic state can be equivalently described by means of the fluid or chain densities of states. The link between the two formulations is the density of states of the Lax matrix $\rho_\mathrm{DOS}(\lambda)$, normalized to one. Since $v=2\lambda$ and $\nu^{-1}=\int dv\rho_\mathrm{qp}(v)$, one has
\be
\rho_{\rm DOS}(\lambda) = 2\nu\rho_\mathrm{qp}(2\lambda)\,.
\ee
Consider now the chain of Sec.~\ref{sec:sr}. In the bulk of the excited region, the Lax operator is the same random Jacobi matrix as $L_N$: the two differ only close to the boundaries, which does not affect the density of states as $N\to\infty$. From the inverse scattering transform, we know that solitons are in one-to-one correspondence with the eigenvalues of the Lax with $|\lambda|>1$. From this, we have the following identification
\be\label{eq:soldens}
\rho_\mathrm{s}(\eta) = |\sinh\eta|\rho_\mathrm{DOS}(\operatorname{sgn}(\eta)\cosh\eta) = 2\nu|\sinh\eta|\rho_\mathrm{qp}(2\operatorname{sgn}(\eta)\cosh\eta)\,.
\ee
On the other side, the identification with the radiation density is more subtle. As we will show in Sec.~\ref{sec:pf}, the radiation density is connected to the Lyapunov exponent $\rho_\mathrm{r}(k)=\frac{-\log|T(e^{ik})|^2}{N\pi}=\tfrac{2}{\pi}\gamma(k)$. This is connected to the density of state of the Lax operator via the Thouless formula~\cite{Thouless1972,TeschlToda}
\be\label{eq:thouless}
\gamma(\lambda) = \log2 +\frac12\nu + \int_\mathbb{R}d\lambda' \log|\lambda-\lambda'|\rho_\mathrm{DOS}(\lambda')\,,
\ee
with $\nu=\langle r_0\rangle$ the average stretch. The rhs shows striking similarities with the TBA expression for the total densities
of the quasiparticles \eqref{eq:qprhot}. Simple manipulations shows the identification $\gamma(\lambda)=\frac{\nu}{2}\rho_\mathrm{qp}^{\bf t}(2\lambda)$. The radiation density is then given by
\be\label{eq:raddens}
\rho_\mathrm{r}(k) = \frac{\nu}{\pi}\rho_\mathrm{qp}^{\bf t}(2\cos k)\, .
\ee
Fig.~\ref{fig:dos} illustrates the identifications \eqref{eq:soldens} and \eqref{eq:raddens} for a thermal state. Note that the factor $|\sinh\eta|$ in \eqref{eq:soldens} makes the soliton density vanish at $\eta=0$: since the width of the soliton \eqref{eq:soliton} is $1/|\eta|$, very extended solitons are not populated.

\begin{figure}[tbp]
    \centering
    \includegraphics[width=\linewidth]{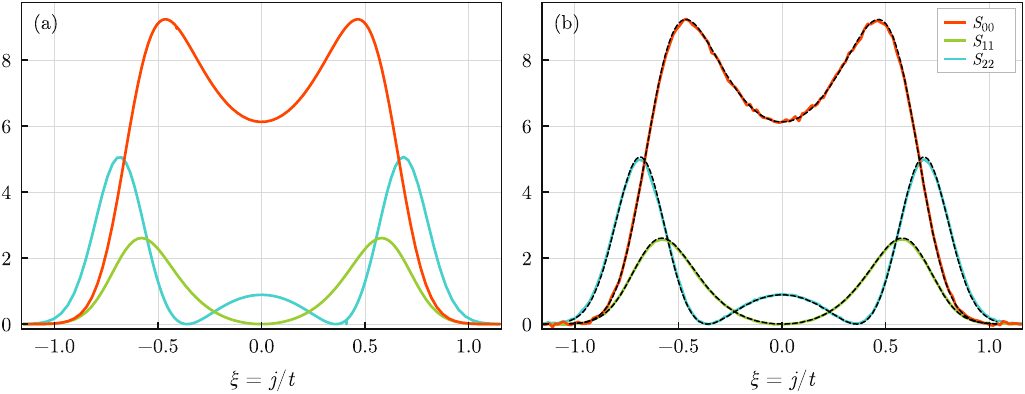}
    \caption{\textbf{Euler-scale dynamical correlators.---} Ballistic scaling profiles $\mathcal{S}_{nn}(\xi)=\lim_{t\to\infty}t\,\langle q_n(\xi t,t)\,q_n(0,0)\rangle_c$ of the stretch ($n=0$), momentum ($n=1$), and energy ($n=2$) connected autocorrelation functions in a thermal state with $\beta=0.5$ and $P=0.32$. The color legend applies to both panels. (a) Mixed soliton--radiation formulation: the densities are obtained by solving the TBA equations~\eqref{eq:TBA}, and the profiles are computed via~\eqref{eq:eulercorr}. (b) Results of Ref.~\cite{mazzuca23}: molecular dynamics data (colored lines) and quasiparticle TBA predictions (black dashed lines). The profiles in (a) coincide with the dashed lines in (b).}
    \label{fig:euler}
\end{figure}

\paragraph{Euler-scale dynamical correlators.}
In Ref.~\cite{mazzuca23}, the Generalized Hydrodynamics equations derived from the fluid TBA were tested against molecular dynamics simulations of a large periodic chain. In particular, the Euler-scale dynamical correlators of the conserved charges, obtained within linear response, were shown to agree with the simulated ones.
To verify our chain TBA equations~\eqref{eq:TBA}, we compute the Euler-scale dynamical correlators of stretch, momentum and energy by solving~\eqref{eq:TBA} and applying the general formula~\eqref{eq:eulercorr}. Fig.~\ref{fig:euler} compares the results for a thermal state at $\beta=0.5$ and $P=0.32$ with those of the fluid TBA and with the molecular dynamics of Ref.~\cite{mazzuca23}. Working Mathematica notebooks for both the chain and fluid TBA are available on Zenodo~\cite{brollo_2026_23696292}.
The agreement is remarkable, since the two formulations are structurally very different: the fluid TBA involves a single type of quasiparticle, whereas ours couples solitons and radiation through three distinct kernels.

%==========================================================
%==========================================================
%==========================================================
\section{Inverse scattering transform}\label{sec:ist}
In this section, I will provide the key ingredients that are needed for the derivation of the free energy functional in Sec.~\ref{sec:pf}. A simple and complete reference, although rather old, is the book by Toda~\cite{Toda81}. 

Most classical integrable systems, both continuous and discrete, admit a Lax pair formulation: the nonlinear equations of motion arise as the compatibility condition of an auxiliary linear problem. The linear problem is defined by two operators, the Lax operator $L$ and the partner operator $B$, together referred to as the \emph{Lax pair}. For Toda, the Lax pair reads
\be
\begin{split}
    &(L f)_j = a_{j} f_{j+1} + \frac12p_j f_j + a_{j-1} f_{j-1}\,, \\
    &(B f)_j =a_{j-1} f_{j-1} - a_{j} f_{j+1}\,,  \\
\end{split}
\ee
where $a_j = \frac12e^{-r_j/2}$ and $f\in\ell^2(\mathbb{Z})$. The Lax matrix $L$ is a symmetric tridiagonal (Jacobi) matrix, while $B$ is antisymmetric.
Both depend on time through $\{a_j(t)\}$ and $\{p_j(t)\}$, which evolve according to \eqref{eq:eomfl}. At any time, $L(t)$ defines a spectral problem, while $B(t)$ generates the time evolution of its eigenvectors,
\be
\begin{split}
    &L(t) \psi(t) = \lambda\psi(t)\,, \\
    &\frac{d}{dt}\psi(t) = B(t)\psi(t)\, .
\end{split}
\ee
The consistency condition between the two is $\dot L=[B,L]$, and it is equivalent to the equations of motion \eqref{eq:eom}. The flow is isospectral, i.e., the spectrum of $L$ is conserved in time.

As in Sec.~\ref{sec:toda}, we consider the infinite chain with initial conditions rapidly decaying to the vacuum as $|j|\to\infty$. Then $a_j \to \frac12$ and $p_j \to 0$, so that $L$ is asymptotically the discrete laplacian, with discrete plane waves solutions. The spectral problem for $L$ is thus a scattering problem on the lattice. This is the setting of the inverse scattering transform. 
By contrast, periodic boundary conditions require a different approach, known as finite-gap integration~\cite{kac1975complete,ferguson1982nonlinear,TeschlToda}, which we will not discuss.

\begin{figure}
    \centering
    \begin{tikzpicture}[node distance=4cm, auto]
      % Nodes
      \node (topL) {$S(L(0))$};
      \node (bottomL) [below of=topL] {$(q_j(0), p_j(0))$};
      \node (topR) [right of=topL] {$S(L(t))$};
      \node (bottomR) [below of=topR] {$(q_j(t), p_j(t))$};
      % Arrows
      \draw[->] (bottomL) -- (topL) node[midway, left] {direct scattering};
      \draw[->] (topL) -- (topR) node[midway, above] {time evolution};
      \draw[->] (topR) -- (bottomR) node[midway, right] {inverse scattering};
    \end{tikzpicture}
    \caption{\textbf{Inverse scattering transform.---} The nonlinear Toda dynamics is linearized by the direct scattering map. Then, the scattering data evolve via \eqref{aatevo}, and the solution at time $t$ is reconstructed by inverse scattering.}
    \label{fig:istscheme}
\end{figure}
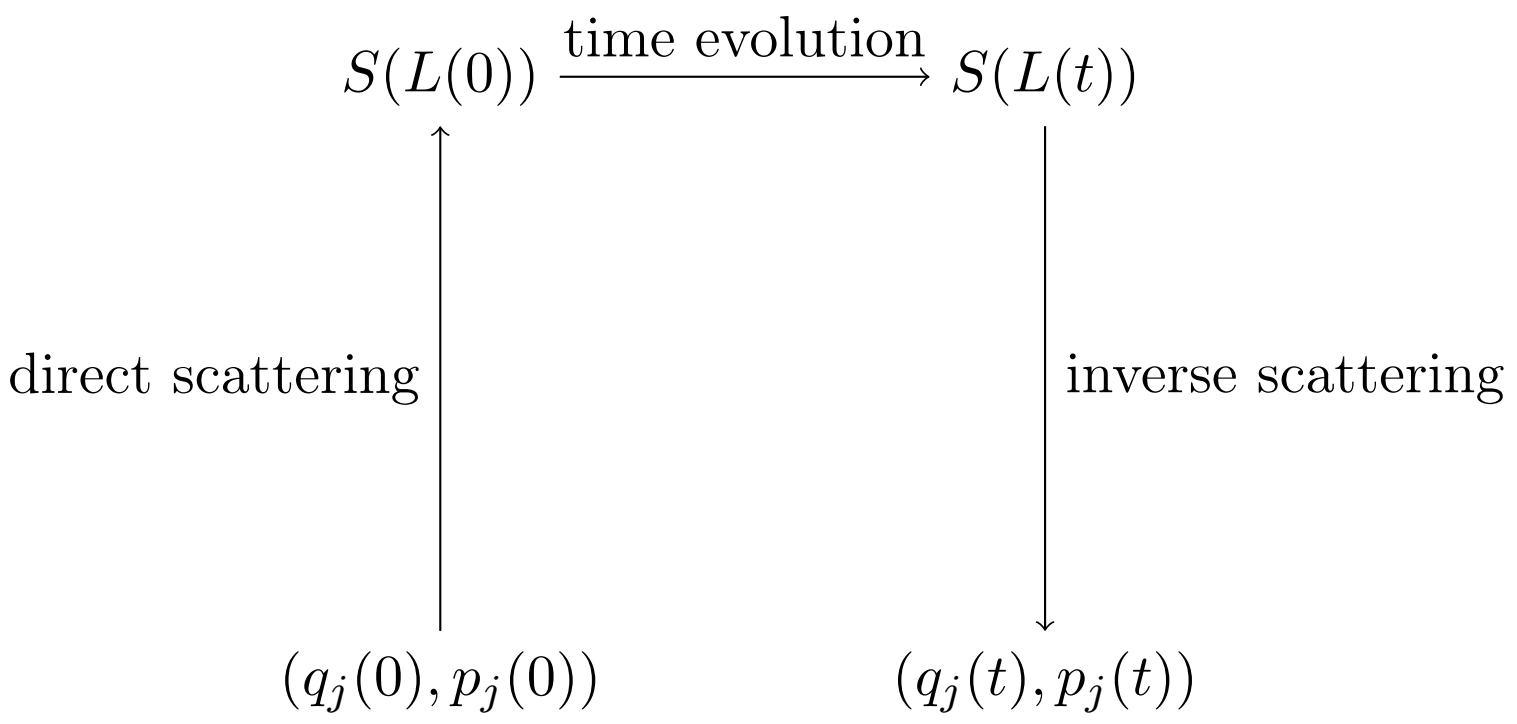

Given initial conditions for the Toda chain at $t=0$, the inverse scattering transform consists of three steps: first, one computes the so-called \emph{scattering data} of the scattering problem defined by $L(t=0)$; these data are then linearly evolved in time by the partner operator $B(t)$; finally, the time evolved Toda coordinates are recovered from the evolved scattering data, either through the Gel'fand-Levitan-Marchenko equation or by solving a Riemann-Hilbert problem. See Fig.~\ref{fig:istscheme}.

\paragraph{Transition matrix and scattering data.} 
From now on, quantities without an explicit time dependence refer to initial-time values.
It is convenient to introduce a new spectral parameter $z \in \mathbb{C}$ via $\lambda = \tfrac{1}{2}(z + z^{-1})$, so the scattering problem reads
\be\label{eq_laxprob}
a_{j-1}\psi_{j-1}(z)
+ \frac{1}{2}p_j\psi_j(z)
+ a_{j}\psi_{j+1}(z)
= \frac{z + z^{-1}}{2}\,\psi_j(z).
\ee
In the asymptotic region $|j| \to \infty$, this reduces to the free tight-binding equation, whose solutions are discrete plane waves. The \emph{Jost solutions} $\varphi_j^\pm(z)$ are the solutions
singled out by the boundary conditions $\varphi_j^\pm(z) \sim z^{\pm j}$ as $j \to \pm\infty$, valid for $|z| \leq 1$. For $|z| = 1$ the solutions are oscillatory and not normalizable, giving rise to the continuous spectrum. The point spectrum lives in the real segment $z\in(-1,0)\cup(0,1)$.

For the continuous spectrum $|z|=1$, the right $(+)$ and left $(-)$ Jost solutions associated to opposite spectral parameters $\varphi_j^+(z),\varphi_j^+(z^{-1})$ and $\varphi_j^-(z),\varphi_j^-(z^{-1})$ are linearly independent, hence defining a basis for all the other solutions. The change-of-basis matrix between the right and left basis defines the \emph{right transition matrix}
\be
\begin{pmatrix}
    \varphi_j^+(z) \\[2pt]
    \varphi_j^+(z^{-1})
\end{pmatrix}
= \mathcal{T}_\mathrm{R}(z)
\begin{pmatrix}
    \varphi_j^-(z^{-1}) \\[2pt]
    \varphi_j^-(z)
\end{pmatrix},
\qquad
\mathcal{T}_\mathrm{R}(z) =
\begin{pmatrix}
    \alpha_\mathrm{R}(z)    & \beta_\mathrm{R}(z) \\
    \beta_\mathrm{R}(z^{-1}) & \alpha_\mathrm{R}(z^{-1})
\end{pmatrix}.
\ee
Analogously, one can define the \emph{left transition matrix} $\mathcal{T}_\mathrm{L}(z)$.
Imposing consistency between the left and right transition matrices forces $\alpha_\mathrm{R}(z) = \alpha_\mathrm{L}(z) \equiv \alpha(z)$, while $\beta_\mathrm{L}(z) = -\beta_\mathrm{R}(z^{-1})$. The entries further satisfy the reality conditions
\be\label{eq_ab}
\bar\alpha(z) = \alpha(\bar z),
\qquad
\overline{\beta}_{\mathrm{R,L}}(z) = \beta_{\mathrm{R,L}}(\bar z),
\qquad
|\alpha(z)|^2 - |\beta_{\mathrm{R,L}}(z)|^2 = 1.
\ee
Since $z^{-1} = \bar z$ on the unit circle, the right transition matrix simplifies to
\be\label{eq_trmat1}
\mathcal{T}_\mathrm{R}(z) =
\begin{pmatrix}
    \alpha(z)           & \beta_\mathrm{R}(z) \\
    \bar\beta_\mathrm{R}(z) & \bar\alpha(z)
\end{pmatrix},
\qquad
\det \mathcal{T}_\mathrm{R}(z) = |\alpha(z)|^2 - |\beta_\mathrm{R}(z)|^2 = 1.
\ee
This matrix satisfies $\mathcal{T}_\mathrm{R}^\dagger(z)\,\sigma_3\,\mathcal{T}_\mathrm{R}(z) = \sigma_3$, with $\sigma_3 = \operatorname{diag}(1,-1)$ the third Pauli matrix, so $\mathcal{T}_\mathrm{R}(z) \in SU(1,1)$.

A physical picture emerges by identifying $\varphi_j^+(z^{-1})$ and $\varphi_j^-(z^{-1})$ as incoming waves, and $\varphi_j^+(z)$ and $\varphi_j^-(z)$ as outgoing waves. One can work in this {in}/{out} basis and define the \emph{scattering matrix}
\be\label{eq:smatrix}
S(z) =
\begin{pmatrix}
    T(z)          & R_\mathrm{L}(z) \\
    R_\mathrm{R}(z) & T(z)
\end{pmatrix},
\qquad |z| = 1,
\ee
where $T(z) = \alpha(z)^{-1}$ is the \emph{transmission coefficient} and $R_{\mathrm{R,L}}(z) = \alpha(z)^{-1}\beta_{\mathrm{R,L}}(z)$ are the right and left \emph{reflection coefficients}. Unitarity of $S(z)$ is equivalent to the on-shell relation $|T(z)|^2 + |R_{\mathrm{R,L}}(z)|^2 = 1$ for $|z| = 1$.  

The transmission amplitude $T(z)$ can be meromorphically extended inside the unit disk, outside the \emph{on-shell} region $|z|=1$. 
Poles occur only for real $z_\alpha \in (-1,0) \cup (0,1)$, they are simple, and the corresponding eigenvalues
$\lambda_\alpha = (z_\alpha + z_\alpha^{-1})/2$ form the
point spectrum $\{\lambda_\alpha\}_{\alpha=1,\ldots,M}$.
The Jost functions corresponding to the poles are normalizable, and we define the right and left \emph{norming constants} $c_{\mathrm{R},\alpha}/c_{\mathrm{L},\alpha}$ as
\be
c_{\mathrm{R},\alpha} = \frac{1}{\sum_j \varphi_j^{+}(z_\alpha)^2}\,,
\qquad c_{\mathrm{L},\alpha} = \frac{1}{\sum_j \varphi_j^{-}(z_\alpha)^2}\,.
\ee
The complete right/left \emph{scattering data} are
\be\label{eq_scattdata}
\mathcal{S}_\mathrm{R,L}
= \bigl\{ R_\mathrm{R,L}(z),\; |z| = 1;\;
  (z_\alpha, c_{\mathrm{R,L},\alpha}),\; 1 \leq \alpha \leq M \bigr\}.
\ee

The transmission coefficient $T(z)$ is determined completely by the scattering data and its analytic structure. Using the Poisson-Jensen formula
\be\label{ist_t}
T(z) =
\left(\prod_{\alpha=1}^{M}
  \frac{|z_\alpha|(z - z_\alpha^{-1})}{z - z_\alpha}
\right)
\exp\!\left(
  \frac{1}{2\pi i}
  \int_{|z_k|=1} \log|T(z_k)|
  \frac{z_k + z}{z_k - z}\,\frac{dz_k}{z_k}
\right),
\qquad |z| \leq 1.
\ee
On the unit circle the integrand is singular, and the integral must
be interpreted via the Sochocki-Plemelj formula
\be\label{spformula}
\frac{1}{z_k - z\,e^{-0}}
= \operatorname{p.v.}\frac{1}{z_k - z}
  + \pi i\,\frac{\delta(z_k z^{-1}-1)}{z},
\ee
where $\operatorname{p.v.}$ denotes the principal value prescription.
Note that for $|z|=1$ only the delta-function term contributes to $|T(z)|$; the principal value part produces a pure phase. Using the reality condition $T(\bar z) = \overline{T(z)}$, the formula can be rewritten as
\be\label{eq_simpt}
T(z) =
\left(\prod_{\alpha=1}^{M}
  \frac{|z_\alpha|(z - z_\alpha^{-1})}{z - z_\alpha}
\right)
\exp\!\left(
  \int_0^\pi dk\;
  \frac{-\log|T(e^{ik})|^2}{\pi}\,
  \frac12\frac{z - z^{-1}}{(z + z^{-1}) - 2\cos k}
\right).
\ee

\paragraph{Lyapunov exponent.}
The spectral problem can equivalently be formulated in terms of a transfer matrix. Rewriting $L\psi(\lambda) = \lambda\psi(\lambda)$ as a recursion, one introduces~\cite{TeschlToda}
\be\label{eq:todaU}
U(\lambda,j) = \frac{1}{a_j}
\begin{pmatrix}
  \lambda - \tfrac{1}{2}p_j & -a_{j-1} \\
  a_{j} & 0
\end{pmatrix},
\ee
so that the eigenvalue equation \eqref{eq_laxprob} is equivalent to
\be\label{eq_transfer}
\begin{pmatrix}
  \psi_{j+1}(\lambda) \\ \psi_j(\lambda)
\end{pmatrix}
= U(\lambda,j)
\begin{pmatrix}
  \psi_j(\lambda) \\ \psi_{j-1}(\lambda)
\end{pmatrix}.
\ee
We define $\Phi(\lambda,j,k)$ the transfer matrix from $k$ to $j$ as 
\be\label{eq:todaPhi}
\Phi(\lambda,j,k) = \overset{\curvearrowleft}{\prod_{l=k}^{j-1}} U(\lambda,l)\,,
\ee
while if $j<k$ it is given by inverse $\Phi(\lambda,j,k)=\Phi(\lambda,k,j)^{-1}$.
The exponential growth rate of a solution is measured by the
\emph{Lyapunov exponent}~\cite{TeschlToda}
\begin{equation}\label{eq_lyap}
\gamma(\lambda)
= \lim_{j \to \infty} \frac{1}{j}
  \log\|\Phi(\lambda,j,0)\|\,,
\end{equation}
with the standard operatorial norm.
For the discrete spectrum, the eigenfunctions decay exponentially in space, which directly implies $\gamma(\lambda) > 0$, while $\gamma(\lambda) = 0$ for the absolutely continuous spectrum, $|\lambda|<1$.

One may ask whether the full transfer
matrix $\lim_{j\to\infty}\Phi(\lambda(z),j,-j)$ is related to the transition matrix $\mathcal{T}_\mathrm{R}(z)$ \eqref{eq_trmat1} for $|z|=1$.
However, the limit does not exist directly, because the asymptotic solutions oscillate. The oscillations are captured by the free transfer matrix $\Phi_0(\lambda,j) =  U_0(\lambda)^j$, where
\be
U_0(\lambda) =
\begin{pmatrix}
  2\lambda & -1 \\ 1 & 0
\end{pmatrix}.
\ee
Diagonalizing $U_0$, we get as eigenvalues $z$ and $z^{-1}$, with $\lambda=\frac12(z+z^{-1})$ (same $z$ defined above)
\be
U_0(\lambda)
= R(z(\lambda))
\begin{pmatrix}
  z(\lambda)^{-1} & 0 \\ 0 & z(\lambda)
\end{pmatrix}
R(z(\lambda))^{-1},
\qquad
R(z) =
\begin{pmatrix}
  1 & -1 \\ z & -z^{-1}
\end{pmatrix},
\ee
the free transfer matrix can be explicitly expressed as
\be
\Phi_0(z,j) = R(z)
\begin{pmatrix}
  z^{-j} & 0 \\ 0 & z^j
\end{pmatrix}
R(z)^{-1}
\ee
One can then strip off the free oscillations from the full transfer matrix, and show that the following limit exists and equals the transition matrix $\mathcal{T}_\mathrm{R}(z)$ in \eqref{eq_trmat1}~\cite{Faddeev87}
\be\label{eq_fulltm}
\lim_{j\to\infty}
R(z)^{-1}\Phi_0(z,-j)\,\Phi(z,j,-j)\,\Phi_0(z,-j)R(z)
=
\begin{pmatrix}
  \alpha(z)           & \beta_\mathrm{R}(z) \\
  \bar\beta_\mathrm{R}(z) & \bar\alpha(z)
\end{pmatrix}\,,\qquad |z|=1\,.
\ee

Let's now connect the transition matrix and the Lyapunov exponent. Consider a situation as in Sec.~\ref{sec:sr}, where the initial conditions in $[0,N]$ are given by a typical configuration sampled from a homogeneous state, such as a GGE, with the chain at rest outside this interval. One can define the Lyapunov exponent for this region, $\gamma_N(\lambda)=N^{-1}\log\|\Phi(\lambda,N,0)\|$, which is self-averaging for large $N$.
Since the chain is free outside $[0,N]$, the limit in \eqref{eq_fulltm} is reached as soon as $j\geq N$. For $\lambda\in(-1,1)$, i.e.\ $|z|=1$, the free factors in \eqref{eq_fulltm} only contribute phases, hence $\gamma_N(\lambda)\simeq N^{-1}\log\|\mathcal{T}_\mathrm{R}(z(\lambda))\|$. One can compute that the operator norm of the right transition matrix is $\|\mathcal{T}_\mathrm{R}(z)\|=|\alpha|+|\beta_\mathrm{R}|$, and since $|\alpha|^2-|\beta_\mathrm{R}|^2=1$, it is of order $|\alpha|$, and therefore $\gamma_N(\lambda)\sim N^{-1}\log|\alpha(z)|$. In terms of the transmission coefficient $T(z)=1/\alpha(z)$, this means $|T(z)|\sim e^{-N\gamma_N(\lambda)}$~\cite{kostrykin99}.

\paragraph{Time evolution and inverse scattering.}
Given the initial data $\{q_j(0),p_j(0)\}$, one computes the right scattering data $S_R(L(0))$~\eqref{eq_scattdata}. The Lax equation implies that they evolve linearly under the partner operator $B$. For the entries of the transition matrix,
\be\label{aatevo}
\dot\alpha(z,t)=0,\qquad
\dot\beta_R(z,t)=-(z-z^{-1})\,\beta_R(z,t),
\ee
and the norming constants evolve as $\beta_R$ evaluated at the poles. The transmission coefficient is hence constant, while the reflection coefficient and the norming constants acquire a time-dependent factor: a pure phase, $z-z^{-1}=i\omega(k)$, for $|z|=1$, and a real exponential at the poles. From \eqref{aatevo} one obtains the linear flow \eqref{eq:pb} of the action-angle variables.

Once the scattering data are evolved to time $t$, the coordinates $(q_j(t),p_j(t))$ are recovered from $S_R(L(t))$. This is the inverse scattering step proper. Classically it is done through the Gel'fand-Levitan-Marchenko equation~\cite{Toda81,flaschka1974toda}, a linear integral equation whose kernel is built from the scattering data. The essential point for us is that the correspondence between the scattering data and the coordinates $\{q_j,p_j\}$ is one-to-one, so that the scattering data are legitimate coordinates on phase space. More recently, the inverse step has been reformulated as a Riemann-Hilbert problem~\cite{kamvissis1993long,Krueger09}. This formulation is suited to the nonlinear steepest descent method of Deift and Zhou~\cite{deift1993steepest}, which gives rigorous long-time asymptotics of the solution, and it also allows for accurate numerics~\cite{Bilman2017}. The asymptotic results obtained in this way for the Toda chain~\cite{Krueger09} are the input of the next section. 

%==========================================================
%==========================================================
%==========================================================
%==========================================================
\section{Soliton and radiation scattering shifts}\label{sec:scattering}

Solitons are known to emerge from mutual collisions unchanged, except for a positional shift given by a sum over pairwise collisions. The reconstruction of the late-time dynamics through the Riemann–Hilbert problem, carried out by Krüger and Teschl~\cite{Krueger09}, shows that the same factorization extends to the scattering with radiation, in the form of an integral over the radiative modes. The radiation behaves in the same way: each radiative mode acquires a phase shift which factorizes into contributions from the solitons and from the other modes, and from this phase one reads off a positional shift.
In this section we recall these results in the form needed in Sec.~\ref{sec:pf}, keeping the notation of Sec.~\ref{sec:toda} and referring to Ref.~\cite{Krueger09} for the derivation.

The key object is the partial transmission coefficient $\tilde T(z)$ of \eqref{eq:pT}
\be
\tilde T(z) =
\left(\prod_{\substack{\alpha=1 \\ z_\alpha \neq z}}^{M}
  \frac{|z_\alpha|(z - z_\alpha^{-1})}{z - z_\alpha}
\right)
\exp\!\left(
  \operatorname{p.v.}\int_0^\pi dk\;
  \frac{-\log|T(e^{ik})|^2}{\pi}\,
  \frac12\frac{z - z^{-1}}{(z + z^{-1}) - 2\cos k}
\right)~.
\ee
It is real for $z$ on the real axis and of unit modulus for $|z|=1$. 
We already discussed that at late times, in the asymptotic regime, a light cone emerges. 
In the region $|j/t|>1$, the solution is a collection of one-solitons \eqref{eq:soliton}, one for each pole $z_\alpha$, with rapidity $\eta_\alpha=-\operatorname{sgn}(z_\alpha)\log|z_\alpha|$ and position parameter
\be\label{eq_solasympt}
\delta^\pm_\alpha=\tilde q_\alpha(0)\pm\log\tilde T(z_\alpha),\qquad t\to\pm\infty,
\ee
where $\tilde q_\alpha(0)=\log|c_{R,\alpha}(0)|$ is the angle variable \eqref{eq:actionangle} at time zero. The peak of the soliton, according \eqref{eq:soliton}, asymptotically moves as
\be\label{eq_solpos}
x_\alpha(t)=v_\mathrm{s}(\eta_\alpha)\,t+q^\pm_\alpha,\qquad
q^\pm_\alpha=\frac{\delta^\pm_\alpha}{2|\eta_\alpha|},\qquad t\to\pm\infty .
\ee
The $q^\pm_\alpha$ are the impact parameters of the soliton, in the sense of Sec.~\ref{sec:cpTBA}: they differ from the rescaled angle $\tilde q_\alpha(0)/2|\eta_\alpha|$ by the shifts accumulated with the other excitations. Comparing future and past,
\be\label{eq_shiftteschl}
q^+_\alpha-q^-_\alpha=\frac{1}{2|\eta_\alpha|}\log\tilde T(z_\alpha)^2=\Delta_\alpha ,
\ee
which is the total shift \eqref{eq:scatt}, and one reads the two-body scattering shifts \eqref{eq:solshifts}.

In the region $|j/t|<1$ the solution is asymptotically dispersive. Along the ray $j/t=\cos k$, $k\in(0,\pi)$,
\be\label{eq_asrad}
a_j(t)=\frac12+\sqrt{\frac{J(k)}{4t}}\,\cos\Big(jk+\frac12\omega(k)\,t+\frac{2J(k)}{\omega(k)}\log t-\delta^\pm(k)\Big)+O(t^{-1}),\qquad t\to\pm\infty,
\ee
with $J(k)$ and $\omega(k)$ as in \eqref{eq:actionangle}. The radiation is thus supported in the light cone, its amplitude decays as $t^{-1/2}$, and each ray isolates a single mode $k$. The asymptotic phases $\delta^\pm(k)$ are given in Ref.~\cite{Krueger09} in terms of the scattering data; what matters here is their structure,
\be\label{eq_radasympt}
\delta^\pm(k)=\frac12\Big[\varphi(k)\mp\arg\tilde T(e^{ik})+\theta\big(J(k)\big)\Big],
\ee
where $\varphi(k)$ is the angle variable \eqref{eq:actionangle} at time zero and $\theta$ is an explicit function of the action $J(k)$ alone. The structure is the same as for the solitons: the phase at time zero, encoded in the angle variable, is shifted by the scattering, in opposite directions in the future and in the past, by an amount depending on the action variables only. Comparing future and past, we have the total phase shift
\be\label{eq_KTphaseshift}
\delta^+(k)-\delta^-(k)=-\arg\tilde T(e^{ik})=i\log\tilde T(e^{ik}).
\ee

The phase, however, is not yet the positional information needed for scattering theory. The radiative mode $k$ is a wavepacket, and, as for a linear dispersive wave, at time $t$ it is located where the phase in \eqref{eq_asrad} is stationary with respect to $k$. This gives
\be\label{eq_radpos}
x_k(t)=v_\mathrm{r}(k)\,t+\xi^\pm(k)+O(\log t),\qquad \xi^\pm(k)=\partial_k\delta^\pm(k),\qquad t\to\pm\infty .
\ee
with $v_\mathrm{r}(k)=\omega'(k)/2=-\cos k$ and the $\xi^\pm(k)$ are the impact parameters of the radiation, the analogue of $q^\pm_\alpha$. Their difference is the scattering shift of the mode,
\be\label{eq_radshift}
\Delta_\mathrm{r}(k)=\xi^+(k)-\xi^-(k)=\partial_k\big[\delta^+(k)-\delta^-(k)\big]=i\,\partial_k\log\tilde T(e^{ik}),
\ee
which is the radiation shift \eqref{eq:scatt}, by which the two-body shifts \eqref{eq:radshifts} are obtained.

%==========================================================
%==========================================================
%==========================================================

\section{Derivation of the chain TBA via inverse scattering transform}\label{sec:pf}
In this section we present the details on how we obtained the free energy functional \eqref{eq:freeE} starting from the microscopic partition function of the Toda chain \eqref{eq:ZL}. The setup is the one of Sec.~\ref{sec:sr}, and we do not repeat it here. We restart from the partition function expressed in terms of the action-angle variables \eqref{eq:ZList},
\be\label{eq:ZLaa}
\begin{split}
&Z_N(\boldsymbol{\beta},P) = \sum_{M}\frac{1}{M!} \int \prod_{\alpha=1}^M d\tilde p_\alpha\exp\left(-\sum_{\alpha=1}^M w_\alpha + P\,\nu_\alpha\right)~\times \\
&\int\mathcal{D} J(k)\exp\left(-\int_0^\pi dk \,A(k)(w(k) + P)\right)\int \prod_{\alpha=1}^M d\tilde q_\alpha\int\mathcal{D}\varphi(k)\,,
\end{split}
\ee
and we refer to that section for the discussion on the extensivity of the radiation.

The section is organized as follows. In Sec.~\ref{sec:cba} we define the functional measure of the radiative sector by discretizing it on the spectrum of the periodic chain. In Sec.~\ref{sec:cpTBA} we recall the derivation of the TBA for classical particles of Ref.~\cite{Doyon19}, since the same steps are reproduced in Sec.~\ref{sec:angles} for the solitons. Finally, in Sec.~\ref{sec:cg} the coarse-graining and the thermodynamic limit lead to the free energy functional \eqref{eq:freeE} and to the coupled TBA equations.

\subsection{Classical Bethe ansatz for radiation}\label{sec:cba}

So far, the radiative part of the partition function has been written as an integral with respect to the formal measure $\mathcal{D}J(k)\,\mathcal{D}\varphi(k)$, without specifying how this measure is defined or on which space of functions it is supported.
The key observation is that the partition function \eqref{eq:ZL} is an integral over a $2N$-dimensional phase space. The map to action--angle variables is canonical and, in particular, preserves this dimension. Since each soliton carries a single action--angle pair $(\tilde p_\alpha,\tilde q_\alpha)$, in the sector with $M$ solitons the radiation data $\big(J(k),\varphi(k)\big)$ live on a $2N_\mathrm{R}$-dimensional space, with $N_\mathrm{R}=N-M$. The radiation can therefore be discretized on $N_\mathrm{R}$ momenta. We will provide an \emph{ansatz} for this discretization based on the spectrum of the periodic problem.

\paragraph{Periodic problem.} Instead of embedding the configuration of the $N$ excited sites into the vacuum, consider it as a stand-alone chain with periodic boundary conditions. In this case, the Lax operator is an $N\times N$ real symmetric matrix, whose $N$ eigenvalues are the roots of $\Delta(\lambda)=2$, where $\Delta=\operatorname{Tr}M_N$ is the discriminant of the monodromy matrix $M_N$ (see App.~\ref{app:periodic}). As shown in App.~\ref{app:periodic}, for $|z|=1$, $M_N$ equals the transition matrix of the embedded configuration, dressed by the free propagation across the interval $[0,N]$. Consequently
\be\label{eq:discr}
\Delta(z)=\alpha(z)\,z^{-N}+\bar\alpha(z)\,z^{N}=2 ,
\ee
where $\alpha(z)$ is the scattering coefficient of \eqref{eq_ab}, related to the transmission coefficient by $T(z)=1/\alpha(z)$.
For the radiative modes, $z=e^{ik}$, the discretization condition \eqref{eq:discr} reads $\cos\!\big(kN+\arg T(e^{ik})\big)=|T(e^{ik})|$. The equation is even in $k$, so it is sufficient to consider it for $k\in(0,\pi)$. 
We obtain the \emph{classical Bethe equations} for the radiation~\cite{de2016equilibration},
\be\label{eq:bethe}
k_nN+\arg T(e^{ik_n})=2\pi n\pm\arccos|T(e^{ik_n})|,
\qquad n\in\mathbb{Z},\quad k_n\in(0,\pi)\,.
\ee
Up to this point the configuration is arbitrary. For a typical configuration of the GGE the transmission is exponentially suppressed, $|T(e^{ik})|\sim\exp(-N\gamma_N(k))$, cfr.~Sec.~\ref{sec:sr}, so the inverse cosine equals $\pi/2$ up to exponentially small corrections, and the Bethe equations reduce to
\be\label{eq:bethe2}
k_nN+\arg T(e^{ik_n})=\pi\Big(n+\frac12\Big),
\qquad n\in\mathbb{Z}\,.
\ee
We then introduce the counting function
\be\label{eq:countingfn}
y(k)=\frac{1}{\pi}\Big(k+\frac{1}{N}\arg T(e^{ik})\Big),
\ee
in terms of which the Bethe equations read $y(k_n)=\big(n+\tfrac12\big)/N$. As $N\to\infty$, the number of roots in $[k,k+dk]$ is $N\,y'(k)\,dk$, i.e.\ $y'(k)$ is the density of radiation momenta,
\be
y'(k)=\frac{1}{\pi}+\frac{1}{\pi N}\,\frac{d}{dk}\arg T(e^{ik}),
\qquad k\in(0,\pi).
\ee
Comparing with the radiation scattering shift \eqref{eq:radshifts}, this becomes
\be\label{eq:radspace}
y'(k)=\frac{1}{\pi}\left(1
-\frac{1}{N}\sum_{\alpha=1}^{M}\Delta_\mathrm{rs}(k,\eta_\alpha)
-\frac{1}{N}\int_0^\pi dk'\,A(k')\,\Delta_\mathrm{rr}(k,k')\right).
\ee
In analogy with quantum integrable systems, where the same Bethe structure appears, we refer to this as the \emph{radiation total density} $\rho_\mathrm{r}^{\bf t}(k)=y'(k)$: it is the density of allowed Bethe roots, and the bare mode spacing $1/\pi$ is dressed by the scattering shifts with all the other excitations. In terms of the kernels \eqref{eq:rhot}, this is the total density of Sec.~\ref{sec:sr}. 
Finally, integrating \eqref{eq:radspace} over $(0,\pi)$ one counts the Bethe roots. An easy computation shows that the number of roots is therefore $N-M=N_\mathrm{R}$, up to $O(1)$ corrections.

\paragraph{Measure.} We denote by $\{k_n\}_{n=1}^{N_\mathrm{R}}$ the set of the Bethe roots of \eqref{eq:bethe2}, and we split $(0,\pi)$ into cells $C_n$, each containing exactly one root, $k_n\in C_n$. Since the roots have density $N\rho_\mathrm{r}^{\bf t}(k)$, the size of the cell $C_n$ is $1/(N\rho_\mathrm{r}^{\bf t}(k_n))$.
The two radiation variables are of different type and must be discretized
differently: $J(k)$ is a \emph{density} (it lives under the $dk$ integral), while $\varphi(k)$ is a \emph{phase} (attached to a point). We therefore integrate the first and sample the second,
\be\label{eq_cellvars}
\mathcal J_n=\int_{C_n}J(k)\,dk,\qquad \varphi_n=\varphi(k_n),
\qquad n=1,\ldots,N_\mathrm{R}.
\ee
Here $\mathcal J_n$ is the action of the single mode in $C_n$. 

We now make the following \emph{ansatz}: in the sector with $M$ solitons, the variables \eqref{eq_cellvars}, together with the solitonic action-angle variables, are good coordinates on the phase space of the $N$ excited sites, and at leading order in $N$ the measure is flat in them, with $(\mathcal J_n,\varphi_n)\in(0,\infty)\times(0,2\pi)$ independently for each mode. We do not have a proof of this statement. We show below that it leads to the TBA equations \eqref{eq:TBA}, whose agreement with the fluid TBA is discussed in Sec.~\ref{sec:results}. Accordingly, we define the radiative measure as
%For the solitons, the requirement that the initial datum be supported on the $N$ excited sites constrains the angle variables, see Sec.~\ref{sec:angles}. For the radiation we assume instead that it is entirely encoded in the choice of the modes, so that the interaction enters the radiative sector only through their density $\rho_\mathrm{r}^{\bf t}$. 
\be\label{eq_discmeasure}
\int \mathcal D J\,\mathcal D\varphi \;=\; \prod_{n=1}^{N_\mathrm{R}} \int_0^\infty d\mathcal J_n\,\int_0^{2\pi}d\varphi_n .
\ee
It is convenient to trade the action $\mathcal J_n$ for the discretized radiation amplitude. Since $J(k)=\tfrac{\omega(k)}{2}A(k)$ and the cells shrink as $N\to\infty$, we set
\be\label{eq_thetan}
\vartheta_n = \int_{C_n}A(k)\,dk = \int_{C_n}\frac{-\log|T(e^{ik})|^2}{\pi}\,dk,
\qquad \mathcal J_n = \frac{\omega(k_n)}{2}\,\vartheta_n ,
\ee
so that $\int_0^\pi dk\, A(k)w(k) = \sum_n\vartheta_n w(k_n)$, provided $w(k)$ is smooth. The change of variables $\mathcal J_n\to\vartheta_n$ produces the Jacobian $\prod_n\omega(k_n)/2$, and the partition function is rewritten as
\be\label{eq:ZLdisc}
\begin{split}
&Z_N(\boldsymbol{\beta},P) = \sum_{M} \frac{1}{M!}
  \int \prod_{\alpha=1}^{M} d\tilde p_\alpha \,
    \exp\biggl[-\sum_{\alpha=1}^{M} \bigl(w_\alpha + P\nu_\alpha\bigr)\biggr]{}\times{} \\
  &\int \prod_{n=1}^{N_\mathrm{R}} d\vartheta_n \,
    \exp\biggl[-\sum_{n=1}^{N_\mathrm{R}} \vartheta_n \bigl(w(k_n) + P\bigr)+\sum_{n=1}^{N_\mathrm{R}}\log\frac{\omega(k_n)}{2}\biggr]
    \int \prod_{\alpha=1}^{M} d\tilde q_\alpha~
    (2\pi)^{N_\mathrm{R}} \,.
\end{split}
\ee
where $(2\pi)^{N_\mathrm{R}}$ comes from the integration of the radiative angle variables, according to the ansatz above.
Note that, since the size of the cell $C_n$ is $1/(N\rho_\mathrm{r}^{\bf t}(k_n))$, for $N\to\infty$ the variables $\vartheta_n$ converge to the continuous function
\be\label{eq_radfilling}
\vartheta_n \longrightarrow \vartheta_\mathrm{r}(k) = \frac{A(k)}{N\rho_\mathrm{r}^{\bf t}(k)}=\frac{\rho_\mathrm{r}(k)}{\rho_\mathrm{r}^{\bf t}(k)},
\ee
with $\rho_\mathrm{r}(k)=A(k)/N$ the radiation density of Sec.~\ref{sec:sr}. This is the \emph{radiative filling function}: it measures the amplitude at which each allowed mode is excited.

\subsection{Derivation of TBA for classical particles}\label{sec:cpTBA}
Our goal is to bring $Z_N(\boldsymbol{\beta},P)$ to a large deviation form, $Z_N(\boldsymbol{\beta},P)\asymp\int\mathcal D\rho\, e^{-N F[\rho]}$, with $F$ a rate functional of the coarse-grained densities of the modes; the free energy is then the minimum of $F$, and the minimizer is the equilibrium state. We report here, schematically, how this is done for a classical integrable gas of particles, following Ref.~\cite{Doyon19}. The same strategy will then be applied to the solitons in Sec.~\ref{sec:angles}.

Consider a classical integrable system, with Hamiltonian $H_\text{cp}(\vec q,\vec p)$ and canonically conjugate variables $(\vec q,\vec p)$. We initially confine the system over a region of volume $R$, we allow the number of particles $N$ to fluctuate, and we sample the initial conditions from a GGE defined by the weight function $W_\text{cp}(\vec q,\vec p)$. The grand canonical partition function is
\be\label{eq_Zcp}
Z_{\text{cp},R} = \sum_{N=0}^\infty\frac{1}{N!}\int\prod_{j=1}^N dp_jdq_j \exp(-W_\text{cp}(\vec q,\vec p) + \mu N)
\ee
where $\mu$ is the chemical potential and the subscript ``cp'' stands for ``classical particles''.
To evaluate this integral, it is convenient to move to the \emph{scattering coordinates}. In integrable systems, as $t\to+\infty$ each particle follows asymptotically free motion,
\be\label{eq_asymptraj}
q_j(t) \approx q_j(0) + \tilde p_j t + \sum_{k:q_k(t)<q_j(t)}\Delta(\tilde p_j,\tilde p_k) - \sum_{k:q_k(0)<q_j(0)}\Delta(\tilde p_j,\tilde p_k)
\ee
where $\tilde p_j=\lim_{t\to+\infty}p_j(t)$ is the asymptotic momentum and $\Delta(\tilde p_j,\tilde p_k)$ is the two-body positional shift acquired during a collision. The first two terms on the right-hand side are the initial position and the asymptotic inertial motion. The last two account for all scattering shifts accumulated between $t=0$ and $t\to\infty$. Due to Liouville's theorem, the \emph{impact parameters}
\be\label{eq_impactparam}
q_j^+ = q_j(0) + \sum_{k:q_k(t)<q_j(t)}\Delta(\tilde p_j,\tilde p_k) - \sum_{k:q_k(0)<q_j(0)}\Delta(\tilde p_j,\tilde p_k)
\ee
together with the asymptotic momenta $\tilde p_j$ form a set of canonically conjugate variables.

The GGE weight $W_\text{cp}(\vec q,\vec p)$ is a function of the conserved charges. In the asymptotic scattering region, where particles no longer interact, it depends only on the asymptotic momenta, $W_\text{cp} = \sum_j w_\text{cp}(\tilde p_j)$. We can now quasi-factorize the partition function, exactly as we did in \eqref{eq:ZList},
\be\label{eq_cpfactpz}
Z_{\text{cp},R} = \sum_{N=0}^\infty\frac{1}{N!}\int_{-\infty}^\infty \prod_{j=1}^N d\tilde p_j\,e^{-\sum_j \left(w_\text{cp}(\tilde p_j)-\mu \right)} \int \prod_{j=1}^N dq^+_j .
\ee

We now evaluate the integral over the impact parameters. Consider the integral in the first variable, $q_1^+$. This integration has a clear geometrical meaning: it counts the effective phase-space volume available to particle 1, given that it must be in the interval $[0,R]$ at $t=0$, and accounting for the scattering shifts due to all other particles. The integration over the other $N-1$ impact parameters can be seen as a measure for the integration in $dq_1^+$, so that particle 1 effectively moves inside a homogeneous bath provided by the other particles. 
Considering the two extreme initial positions of particle 1, the leftmost $q_1(0)=0$, where it scatters with all slower particles ($\tilde p_k<\tilde p_1$), and the rightmost $q_1(0)=R$, where it scatters with all the faster ones ($\tilde p_k>\tilde p_1$),  we obtain
\be\label{eq_intip}
\int dq^+_1 = \left(R - \sum_{\tilde p_k> \tilde p_1}\Delta(\tilde p_1, \tilde p_k)\right) - \left( \sum_{\tilde p_k< \tilde p_1}\Delta(\tilde p_1, \tilde p_k)\right) = R\left(1 - \frac{1}{R}\sum_{k\neq 1}\Delta(\tilde p_1, \tilde p_k) \right) ,
\ee
in the sign convention for the shifts of \eqref{eq:scatt}.
The same result holds for every particle. Substituting back into \eqref{eq_cpfactpz},
\be\label{eq_zcpfiniteL}
Z_{\text{cp},R} = \sum_{N=0}^\infty\frac{R^N}{N!}\int_{-\infty}^\infty \prod_{j=1}^N d\tilde p_j \exp\Big[-\sum_{j=1}^N \Big(w_\text{cp}(\tilde p_j)-\mu - \log\Big(1 - \frac{1}{R}\sum_{k\neq j}\Delta(\tilde p_j, \tilde p_k)\Big)\Big) \Big] .
\ee
The bare volume $R$ available to each particle is renormalized by the scattering shifts with all the others. This is the only place where the interaction enters.

We now take the thermodynamic limit $R\to\infty$. Let $\nu=R/N$ denote the mean volume per particle and introduce the empirical density
\be
\tilde\rho_N(p) = \frac{1}{N}\sum_{j=1}^N\delta(p-\tilde p_j),
\ee
which converges, as $N\to\infty$, to a smooth density $\tilde\rho(p)$, normalized to unity. The integral over the momenta is coarse-grained accordingly: the product measure is replaced by $\mathcal{D}\tilde\rho(p)\,e^{NS[\tilde\rho]}$, where $S[\tilde\rho]=-\int dp\,\tilde\rho(p)\log\tilde\rho(p)$ is the Shannon entropy, which counts the number of momentum configurations compatible with a given $\tilde\rho$. Together with the Stirling approximation $R^N/N!\approx \exp(N\log\nu+N)$, the partition function takes the large deviation form
\be\label{eq_LDcp}
Z_{\text{cp},R}\asymp \int\mathcal D\rho\; e^{-R\,F_\text{cp}[\rho]},
\ee
where, rescaling to the particle density per unit volume $\rho(p)=\nu^{-1}\tilde\rho(p)$, the free energy functional reads
\be\label{eq_Fcp}
F_\text{cp}[\rho] = \int dp\,\rho(p)\left(w_\text{cp}(p)-\mu - \log\rho^{\bf t}(p) + \log\rho(p) - 1 \right),
\qquad \rho^{\bf t}(p) = 1-\int dp'\Delta(p,p')\rho(p').
\ee
Here $\rho^{\bf t}$ is the total density of states, in the standard TBA notation of Sec.~\ref{sec:tba}. Note that $\rho$ is not normalized, since $N$ fluctuates, and the sum over $N$ is included in the functional integral. The free energy density is the minimum of the rate functional,
\be
f_\text{cp}(\boldsymbol\beta,\mu) = -\lim_{R\to\infty}\frac1R\log Z_{\text{cp},R} = \min_{\rho}F_\text{cp}[\rho],
\ee
and the equilibrium density is the minimizer. Introducing the filling function $\vartheta(p)=\rho(p)/\rho^{\bf t}(p)$ and the pseudoenergy $\vartheta(p)=\exp(-\varepsilon(p))$, the stationarity condition $\delta F_\text{cp}/\delta \rho = 0$ yields the classical TBA equation
\be\label{eq_TBAcp}
\varepsilon(p) = w_\text{cp}(p) - \mu + \int dp'\Delta(p,p')\exp(-\varepsilon(p')) .
\ee
Comparing \eqref{eq_Fcp} with the general form of Sec.~\ref{sec:tba}, the entropy density is $s_\text{cp}[\rho]=-\int dp\,\rho(p)(\log\vartheta(p)-1)$, the Maxwell--Boltzmann functional. It arises from two distinct contributions: the integral over the angle variables, which gives the renormalized phase-space volume available to each mode ($\log\rho^{\bf t}$), and the coarse-graining of the momentum distribution ($-\log\rho+1$). We will see that for radiation the two contributions are organized differently.

\subsection{Integration over the angle variables}\label{sec:angles}
We now integrate the solitonic angle variables in \eqref{eq:ZLdisc}, following the steps that led from \eqref{eq_cpfactpz} to \eqref{eq_zcpfiniteL}.
Solitons are in close analogy with classical particles. In Sec.~\ref{sec:toda} we parametrized them by the rapidity $\eta\in\mathbb{R}$, in one-to-one correspondence with the action variable via $\tilde p_\alpha = -\operatorname{sgn}(\eta_\alpha)\,2\cosh\eta_\alpha$, and we accordingly rescale the angle variable, defining $q_\alpha = \tilde q_\alpha/2|\eta_\alpha|$. As one reads from \eqref{eq_solasympt} and \eqref{eq_solpos}, the rescaled angles $q_\alpha$ differ from the impact parameters $q_\alpha^+$ only by scattering shifts, that by functions of the action variables only. The map from $q_\alpha$ to $q_\alpha^+$ is in general not canonical~\cite{Faddeev87}, but for this very reason the change of variables $\{\eta_\alpha, q_\alpha\}\to\{\eta_\alpha, q_\alpha^+\}$ has unit Jacobian determinant, and we can perform it freely. In these variables the partition function reads
\be\label{eq_Zsolip}
\begin{split}
Z_N(\boldsymbol{\beta},P) &= \sum_M\frac{1}{M!} \int \prod_{\alpha=1}^M d\eta_\alpha
\exp\Big(-\sum_{\alpha=1}^M w(\eta_\alpha) + P\sum_{\alpha=1}^M 2|\eta_\alpha|
+ \sum_{\alpha=1}^M \log|2\sinh\eta_\alpha|\Big)~\times \\
&\int \prod_{n=1}^{N_\mathrm{R}} d\vartheta_n\exp\Big(-\sum_{n=1}^{N_\mathrm{R}}\vartheta_n\big(w(k_n)+P\big)+\sum_{n=1}^{N_\mathrm{R}}\log\pi\omega(k_n)\Big)
\int \prod_{\alpha=1}^M 2|\eta_\alpha|\, dq_\alpha^+,
\end{split}
\ee
where we substituted the soliton stretch $\nu_\alpha=-2|\eta_\alpha|$, the term $\sum_\alpha \log|2\sinh\eta_\alpha|$ in the exponent is the Jacobian $|d\tilde p_\alpha/d\eta_\alpha|=2\sinh|\eta_\alpha|$ of the change of variables $\tilde p_\alpha \to \eta_\alpha$, and the factor $\prod_\alpha 2|\eta_\alpha|$ is the Jacobian of the rescaling of the angles.

The partition function is now in the same form as \eqref{eq_cpfactpz}, and the integration over $dq_\alpha^+$ is performed exactly as in the classical particle case \eqref{eq_intip}: the bare phase space volume of a soliton, here played by the number of sites $N$, is renormalized by the scattering shifts \eqref{eq:solshifts} with all the other solitons and with the radiation. The only difference is that the mode integral over the radiation in \eqref{eq:scatt} is replaced by a sum over the Bethe modes, weighted by the amplitudes $\vartheta_n$,
\be\label{eq_solvolume}
2|\eta_\alpha|\int dq_\alpha^+
= N\left(2|\eta_\alpha|
- \frac{1}{N}\sum_{\substack{m=1 \\ m\neq \alpha}}^{M}T_\text{ss}(\eta_\alpha,\eta_m)
- \frac{1}{N}\sum_{n=1}^{N_\mathrm{R}} \vartheta_n\,T_\text{sr}(\eta_\alpha,k_n)\right)
\xrightarrow[N\to\infty]{} N\rho_\mathrm{s}^{\bf t}(\eta_\alpha) ,
\ee
where $T_\mathrm{ss}(\eta,\eta')=2|\eta|\Delta_\mathrm{ss}(\eta,\eta')$ and $T_\mathrm{sr}(\eta,k)=2|\eta|\Delta_\mathrm{sr}(\eta,k)$ are the soliton kernels, and the right-hand side converges to the soliton total density $\rho_\mathrm{s}^{\bf t}$ \eqref{eq:spacedensity}. Substituting back into the partition function, we arrive at
\be\label{eq_Zangint}
\begin{split}
Z_N(\boldsymbol{\beta},P) = &\sum_M\frac{N^M}{M!} \int \prod_{\alpha=1}^M d\eta_\alpha\,
\exp\Big(-\sum_{\alpha=1}^M w(\eta_\alpha) + P\sum_{\alpha=1}^M 2|\eta_\alpha|
+ \sum_{\alpha=1}^M \log|2\sinh\eta_\alpha| + \sum_{\alpha=1}^M \log\rho_\mathrm{s}^{\bf t}(\eta_\alpha)\Big)~\times \\
&\int \prod_{n=1}^{N_\mathrm{R}} d\vartheta_n\exp\Big(-\sum_{n=1}^{N_\mathrm{R}}\vartheta_n\big(w(k_n)+P\big)+\sum_{n=1}^{N_\mathrm{R}}\log\pi\omega(k_n)\Big)\,.
\end{split}
\ee

\subsection{Coarse-graining and thermodynamic limit}\label{sec:cg}
We now take the thermodynamic limit $N\to\infty$ of \eqref{eq_Zangint}, coarse-graining the solitonic and the radiative action variables into the densities $\rho_\mathrm{s}(\eta)$ and $\rho_\mathrm{r}(k)$ of Sec.~\ref{sec:sr}. For the solitons this works exactly as for the classical particles of Sec.~\ref{sec:cpTBA}, while for the radiation the coarse-graining is of different nature and, as we will see, it is at the origin of the Rayleigh-Jeans entropy.

\paragraph{Solitons.} 
We define the empirical density of the soliton rapidities 
\be
\tilde \rho_\mathrm{s}^{[M]}(\eta) = \frac{1}{M}\sum_{\alpha=1}^M \delta(\eta - \eta_\alpha),
\ee
which converges to a smooth density $\tilde\rho_\mathrm{s}(\eta)$, normalized to unity, as $M\to\infty$. As in Sec.~\ref{sec:cpTBA}, the product measure $\prod_\alpha d\eta_\alpha$ is replaced by $\mathcal D\tilde\rho_\mathrm{s}\,e^{MS[\tilde\rho_\mathrm{s}]}$, with $S[\tilde\rho_\mathrm{s}]=-\int d\eta\,\tilde\rho_\mathrm{s}(\eta)\log\tilde\rho_\mathrm{s}(\eta)$ the Shannon entropy, and the prefactor is approximated as $N^M/M!\approx \exp(M\log\nu_\mathrm{s} + M)$, with $\nu_\mathrm{s}=N/M$ the number of sites per soliton. Rescaling to the soliton density per site, $\rho_\mathrm{s}(\eta)=\nu_\mathrm{s}^{-1}\tilde\rho_\mathrm{s}(\eta)$, the solitonic part of \eqref{eq_Zangint} becomes $\int\mathcal D\rho_\mathrm{s}\,e^{-NF_\mathrm{S}[\rho_\mathrm{s},\rho_\mathrm{r}]}$, with the sum over $M$ absorbed in the functional integral, and
\be\label{eq_FS}
F_\mathrm{S}[\rho_\mathrm{s},\rho_\mathrm{r}] = \int d\eta\,\rho_\mathrm{s}(\eta)
\Big (w(\eta) - 2P|\eta|
- \log|2\sinh\eta| - \log\rho_\mathrm{s}^{\bf t}(\eta) + \log\rho_\mathrm{s}(\eta) - 1\Big).
\ee
The dependence on the radiation enters through the total density $\rho_\mathrm{s}^{\bf t}$, cfr. \eqref{eq:spacedensity}. Up to the term $\log|2\sinh\eta|$, which originates from the parametrization of the solitons by the rapidity, this is the same functional as for the classical particles \eqref{eq_Fcp}: the soliton entropy has the Maxwell-Boltzmann form, and it arises from the positional freedom (the integration over the impact parameters) and from the indistinguishability of the solitons (the coarse-graining of the rapidities, together with the $1/M!$).

\paragraph{Radiation.}
As $N\to\infty$, the number of radiative modes $N_\mathrm{R}\to\infty$ as well, and the discrete amplitudes $\vartheta_n$ are coarse-grained into the continuous filling function $\vartheta_\mathrm{r}(k)$ in \eqref{eq_radfilling}. Consider the modes with $k_n\in(k,k+dk)$: their number is $N_k=N\rho_\mathrm{r}^{\bf t}(k)dk$, and their amplitudes are summed up into
\be
\vartheta_\mathrm{r}(k) = \frac{1}{N_k}\sum_{n:\,k_n\in(k,k+dk)} \vartheta_n .
\ee
In contrast with the soliton case, here the positions of the modes are fixed (by the Bethe equations), and it is the amplitudes that fluctuate: the coarse-graining counts the number of ways of distributing the total amplitude of the window among its $N_k$ modes. This is the microcanonical integral
\be\label{eq_simplex}
\int_0^\infty \prod_{n=1}^{N_k} d\vartheta_n\;\delta\Big(\sum_{n=1}^{N_k}\vartheta_n - N_k\,\vartheta_\mathrm{r}\Big)
= \frac{(N_k\,\vartheta_\mathrm{r})^{N_k-1}}{(N_k-1)!}
\asymp \exp\big[N_k\big(\log\vartheta_\mathrm{r}+1\big)\big]\,.
\ee
Multiplying over all the windows, the product measure is replaced by
\be\label{eq_RJmeasure}
\prod_{n=1}^{N_\mathrm{R}} d\vartheta_n
= \mathcal{D}\vartheta_\mathrm{r}(k)\,
\exp\Big(N\int_0^\pi dk\,\rho_\mathrm{r}^{\bf t}(k)\big(\log\vartheta_\mathrm{r}(k)+1\big)\Big).
\ee
This is the Rayleigh--Jeans entropy of a classical field, and it is the radiative counterpart of the Shannon entropy of the solitons. Note the different organization of the two contributions discussed at the end of Sec.~\ref{sec:cpTBA}: for radiation, by the ansatz of Sec.~\ref{sec:cba}, the integration over the angle variables gives no renormalized volume: the interaction enters through the density of the modes, and the entropy is entirely due to the coarse-graining of the amplitude.

Finally, the remaining sums over the modes in \eqref{eq_Zangint} are converted into integrals, $\sum_n g(k_n)\to N\int_0^\pi dk\,\rho_\mathrm{r}^{\bf t}(k)g(k)$, and the radiative part of the partition function takes the form $\int\mathcal D\vartheta_\mathrm{r}\,e^{-NF_\mathrm{R}[\rho_\mathrm{s},\rho_\mathrm{r}]}$ with
\be\label{eq_FR}
F_\mathrm{R}[\rho_\mathrm{s},\rho_\mathrm{r}] = \int_0^\pi dk\,\rho_\mathrm{r}^{\bf t}(k)
\Big[\big(w(k)+P\big)\vartheta_\mathrm{r}(k) - \log\big(\pi\omega(k)\vartheta_\mathrm{r}(k)\big) - 1\Big],
\ee
where the dependence on the solitons enters through $\rho_\mathrm{r}^{\bf t}$, cfr. \eqref{eq:spacedensity}, and the term $\log(\pi\omega(k))$ is the Jacobian of the passage from the action $\mathcal J_n$ to the amplitude $\vartheta_n$ in \eqref{eq_thetan}, combined with the $2\pi$ of the phase integration.

\paragraph{Large deviation form and coupled TBA equations.}
Collecting the two sectors, the partition function takes the large deviation form
\be\label{eq_LDtoda}
Z_N(\boldsymbol\beta,P) \asymp \int \mathcal{D}\rho_\mathrm{s}(\eta)\,\mathcal{D}\rho_\mathrm{r}(k)\;
e^{-N\big(F_\mathrm{S}[\rho_\mathrm{s},\rho_\mathrm{r}] + F_\mathrm{R}[\rho_\mathrm{s},\rho_\mathrm{r}]\big)},
\ee
where we traded $\vartheta_\mathrm{r}$ for the radiation density $\rho_\mathrm{r}=\rho_\mathrm{r}^{\bf t}\vartheta_\mathrm{r}$, a change of variables that is irrelevant for the stationarity conditions. The rate functionals $F_\mathrm{S}$ and $F_\mathrm{R}$ are the free energy functionals \eqref{eq:freeE}. The two sectors are coupled only through the total densities. The free energy per site is the minimum of the rate functional,
\be\label{eq_fmin}
f(\boldsymbol\beta,P) = -\lim_{N\to\infty}\frac{1}{N}\log Z_N(\boldsymbol\beta,P) = \min_{\rho_\mathrm{s},\,\rho_\mathrm{r}}\big(F_\mathrm{S} + F_\mathrm{R}\big),
\ee
and the equilibrium densities are the minimizers. Both sectors are grand canonical, so the variations are free. Parametrizing the fillings by the pseudoenergies,
\be\label{eq_statistics}
\vartheta_\mathrm{s}(\eta) = e^{-\varepsilon_\mathrm{s}(\eta)},
\qquad
\vartheta_\mathrm{r}(k) = \frac{1}{\varepsilon_\mathrm{r}(k)},
\ee
i.e. Maxwell-Boltzmann statistics for the solitons and Rayleigh--Jeans statistics for the radiation, the stationarity conditions $\delta(F_\mathrm{S}+F_\mathrm{R})/\delta\rho_\mathrm{s}=\delta(F_\mathrm{S}+F_\mathrm{R})/\delta\rho_\mathrm{r}=0$ give the coupled TBA equations
\be\label{eq_TBA}
\begin{split}
\varepsilon_\mathrm{s}(\eta) &= w(\eta) - 2P|\eta| - \log|2\sinh\eta|
+ \int d\eta'\,T_\text{ss}(\eta,\eta')\,e^{-\varepsilon_\mathrm{s}(\eta')}
+ \int_0^\pi dk\,T_\text{rs}(k,\eta)\,\log\frac{\varepsilon_\mathrm{r}(k)}{\pi\omega(k)},\\[1mm]
\varepsilon_\mathrm{r}(k) &= w(k) + P
+ \int d\eta\,T_\text{sr}(\eta,k)\,e^{-\varepsilon_\mathrm{s}(\eta)}
+ \int_0^\pi dk'\,T_\text{rr}(k',k)\,\log\frac{\varepsilon_\mathrm{r}(k')}{\pi\omega(k')},
\end{split}
\ee
with the kernels of \eqref{eq:kernels}. These are the TBA equations \eqref{eq:TBA} of Sec.~\ref{sec:sr}. This concludes the derivation, which for the radiative sector relies on the ansatz of Sec.~\ref{sec:cba}.

%==========================================================
%==========================================================
%==========================================================
\section{Other models and limitations}\label{sec:others}
We now discuss how we expect the procedure developed above to apply to other models solvable by the inverse scattering transform.
The discussion is kept at a general level, and we leave detailed results to future work. We focus on models whose TBA has already been discussed in the literature, in order to highlight analogies and differences with our approach:
the Korteweg de Vries equation~\cite{bonnemain22}, the nonlinear Schr\"odinger equations~\cite{del2020exact,koch22} and both the sinh- and sine-Gordon models~\cite{de2016equilibration,Koch23}.

\subsection{Korteweg de Vries equation}
The KdV equation is an integrable nonlinear PDE for the real field $u(x,t)$
\be
\partial_t u + 6u\partial_x u + \partial^3_{xxx} u = 0\,.
\ee
As for Toda, we consider the system on the whole line with rapidly decaying boundary conditions $u(x)\to0$ as $|x|\to\infty$.
The equations can be rewritten in Lax form, with Lax operator $\mathcal{L}=-\partial^2_x+u(x,t)$ and appropriate partner operator. The scattering problem induced by the Lax operator is given by a Schr\"odinger equation with the initial time field $u(x)$ as potential~\cite{gardner1967method}
\be
-\p_x^2 \psi(x;k) + u(x) \psi(x;k) = k^2 \psi(x;k)\,. 
\ee
The spectrum is absolutely continuous for $k\in\mathbb{R}$ and discrete along the positive imaginary axis, with eigenvalues $i\kappa_\alpha$, $\kappa_\alpha>0$ and $\alpha=1,\ldots,M$. The spectrum of the Lax operator is conserved by the time evolution.
The transmission coefficient is meromorphic on the upper half plane $\operatorname{Im}k>0$, with simple poles in correspondence of $i\kappa_\alpha$. It takes the form~\cite{deift1979inverse}
\be\label{eq:kdvT}
T(k) = \prod_{\alpha=1}^M \frac{k+i\kappa_\alpha}{k-i\kappa_\alpha}\exp\Big(-\frac{ik}{\pi}\int_0^\infty dq \frac{\log|T(q)|^2}{q^2-k^2}\,\Big)\,\qquad \operatorname{Im}k>0\,,
\ee
and on the real axis is given by the prescription $k+i0^+$. 

The long-time asymptotics for rapidly decaying initial data have been derived
rigorously via the nonlinear steepest descent method~\cite{deift1994collisionless,grunert2009long}.
As for the Toda lattice, two main regions emerge. For $x/t<0$ the solution
consists of dispersive radiation decaying as $t^{-1/2}$, whose amplitude is
controlled by $A(k)=-\tfrac{1}{\pi}\log|T(k)|^2$. For $x/t>0$ it
asymptotically reduces to a train of solitons, one for each pole of $T(k)$.
The scattering shifts can be extracted from the rigorous asymptotics~\cite{grunert2009long}, following considerations analogous to those of Sec.~\ref{sec:scattering}. The shifts among solitons (ss), among radiation modes (rr), and between solitons and radiation (sr, rs) are 
\be
\Delta_{\rm ss}(\kappa,\kappa')
    =\frac{1}{\kappa}\log\Big|\frac{\kappa+\kappa'}{\kappa-\kappa'}\Big|\,,
\qquad
\Delta_{\rm sr}(\kappa,k)
=-\frac{1}{\kappa^2+k^2}\,,
\ee
\be
\Delta_{\rm rr}(k,k')
    \;=\;\frac{d}{d k}\,\text{p.v.}\,\frac{k}{k^2-k'^2}\, ,
\qquad
\Delta_{\rm rs}(k,\kappa)
\;=\;\frac{2\kappa}{\kappa^2+k^2}\, .
\ee
The local conserved charges follow from the large-$k$ expansion of the transmission coefficient, finding~\cite{zakharov1971korteweg}
\be
Q_n=\sum_{\alpha=1}^{M}\frac{2\,\kappa_\alpha^{2n+1}}{2n+1}
      +(-1)^{n+1}\int_0^\infty\!dk\,A(k)\,k^{2n}\,,
\label{eq:kdvcharges}
\ee
normalized such that $Q_0=\tfrac12\int dx\,u$, $Q_1=\tfrac18\int dx\,u^2$ and $Q_2=-H/16$.

The KdV equation admits a Hamiltonian form~\cite{NMPZ84}
\be
\partial_t u=\{u,H\}=\partial_x\frac{\delta H}{\delta u}\,,\qquad
H=\int dx\,\Big(\tfrac12(\partial_x u)^2-u^3\Big),
\ee
with respect to the non-canonical bracket
\be
\{F,G\}=\int dx\,\frac{\delta F}{\delta u}\,\partial_x\frac{\delta G}{\delta u}\,,
\qquad\text{i.e.}\qquad
\{u(x),u(y)\}=\delta'(x-y)\,.
\ee
As in \eqref{eq:actionangle}, the inverse scattering transform provides action-angle variables in terms of the scattering data
\be
P(k)=-\frac{2k}{\pi}\,\log |T(k)|\,,\qquad \theta(k)=\arg\frac{R(k)}{T(k)}\,,
\ee
\be
p_\alpha=\kappa_\alpha^2\,,\qquad \theta_\alpha=-2\log|c_\alpha|\,,
\ee
where $R(k)$ and $c_\alpha$ are the right reflection coefficient and norming constants, respectively. They obey canonical brackets
\be
\{\theta(k),P(k')\}=\delta(k-k')\,,\qquad \{\theta_\alpha,p_\beta\}=\delta_{\alpha\beta}\,,
\ee
while all other brackets vanishing. The dynamics is then linear, $\theta(k,t)=\theta(k,0)+8k^3t$ and $\theta_\alpha(t)=\theta_\alpha(0)-16\kappa_\alpha^3t$.

\paragraph{Thermodynamic Bethe Ansatz.}
As in Sec.~\ref{sec:sr}, we take the field to be in its vacuum outside an interval of length $\ell$. Inside the interval it is distributed as $e^{-W}$, $W=\sum_n\beta_nQ_n$, with respect to the flat measure $\mathcal{D}u(x)=\prod_x du(x)$. This measure is only formal and the what are the appropriate $W$ that gives a correct thermodynamics ha to be discuss. However, we will not address this problems here, and we assume everything is fine, as it is in Toda.

Under these assumptions, we proceed changing variables to the action-angle ones, and by~\eqref{eq:kdvcharges}, $W=\sum_\alpha w_s(\kappa_\alpha)+\int_0^\infty dk\,A(k)\,w_r(k)$ with
\begin{equation}
  w_s(\kappa)=\sum_n\beta_n\frac{2\,\kappa^{2n+1}}{2n+1},\qquad
  w_r(k)=\sum_n\beta_n(-1)^{n+1}k^{2n}.
  \label{eq:kdvgge}
\end{equation}
The result is similar to \eqref{eq:ZList}.
Again, as $\ell\to\infty$ we expect the mean soliton number grows extensively, while the extensivity of radiation comes from localization, and in here it is literally Anderson localization. For $u$ random, stationary and ergodic, the scattering problem is a 1D Schr\"odinger equation in a random potential, whose Lyapunov exponent $\gamma(k)$ is positive for almost every $k$. Hence $|T(k)|\sim e^{-\ell\gamma_\ell(k)}$, leading to extensivity of radiation. 

We then expect the derivation of Sec.~\ref{sec:pf} to extend to KdV, with possible additional complications due to the ultraviolet divergences typical of classical field theories.
We expect the final result to be a coupled soliton-radiation TBA, in close analogy with~\eqref{eq:TBA}, with Boltzmann statistics for the solitons and Rayleigh-Jeans statistics for the radiation. This contrasts with Ref.~\cite{bonnemain22}, where a purely solitonic TBA is constructed and argued as a complete description of KdV thermodynamics. Further work is needed to clarify this point.

\subsection{Nonlinear Schr\"odinger, sinh-Gordon and sine-Gordon equations}
Relevant integrable nonlinear PDEs for their wide applicability in physics are the nonlinear Schr\"odinger equations for the complex field $\psi(x)$
\be
i\partial_t \psi(x,t) = -\partial_x^2 \psi(x,t) + 2\sigma\,|\psi(x,t)|^2\psi(x,t),
\ee
where $\sigma=1$ ($\sigma=-1$) corresponds to the defocusing (focusing) case, and the sinh- and sine-Gordon equations for the real field $\phi(x)$
\be
\partial_t^2\phi - \partial_x^2\phi + \sinh\phi = 0, \qquad
\partial_t^2\phi - \partial_x^2\phi + \sin\phi = 0 .
\ee
They are also solved via the inverse scattering transform, but the scattering problem is now two-dimensional and given by the Zakharov-Shabat problem~\cite{Ablowitz2003}
\be\label{eq:zs}
\begin{pmatrix} i\partial_x & -i\,q(x) \\ i\,r(x) & -i\partial_x \end{pmatrix}
\begin{pmatrix} \chi_1 \\ \chi_2 \end{pmatrix}
= \lambda \begin{pmatrix} \chi_1 \\ \chi_2 \end{pmatrix}\,
\ee
with spectral parameter $\lambda\in\mathbb{C}$ and the potentials $q,r$ are fixed by the initial data. From now on, quantities without an explicit time dependence refer to initial-time values. For the nonlinear Schr\"odinger equation the connection is direct~\cite{Ablowitz2003}: $q(x)=\psi(x)$ and $r(x)=\sigma\,\psi^*(x)$. For $\sigma=1$ the operator in \eqref{eq:zs} is self-adjoint and, for decaying initial data its spectrum is purely absolutely continuous and lies on the real axis, meaning  the defocusing case has radiation as its only type of excitation. 
Conversely, for $\sigma=-1$, the operator is not self-adjoint and discrete eigenvalues $\lambda_j\in\mathbb{C}\setminus\mathbb{R}$ may appear, meaning that the focusing equation supports solitons in addition to radiation. 

Equivalently, \eqref{eq:zs} can be written as
\be\label{eq:zs1}
\partial_x \chi = U_\pm(\lambda) \chi\,,\qquad\text{with}\qquad
U_\pm(\lambda) = 
\begin{pmatrix} -i\lambda & \psi(x) \\ \pm\psi^*(x) & i\lambda \end{pmatrix}\,,
\ee
so that the transfer matrix from $y$ to $x$ is the path-ordered exponential
\be
T(\lambda;x,y)=\mathcal{P}\exp\int_y^x\! dx'\,U_\pm(\lambda)\,,
\ee
and, as in \eqref{eq_fulltm}, the transition matrix is obtained by stripping off the free oscillations~\cite{Faddeev87},
\be
\mathcal{T}(\lambda)=\lim_{x\to\infty,\;y\to-\infty} e^{i\lambda\sigma_3 x}\,T(\lambda;x,y)\,e^{-i\lambda\sigma_3 y}\,.
\ee
In the defocusing case, $U_+(\lambda)$ for $\lambda\in\mathbb{R}$ is traceless and satisfies $U_+^\dagger\sigma_3+\sigma_3 U_+=0$, i.e.\ it belongs to the Lie algebra $\mathfrak{su}(1,1)$, and hence $T(\lambda;x,y)$, as well as $\mathcal{T}(\lambda)$, belongs to $SU(1,1)$. This group is non-compact, so that for appropriately random initial data one expects a positive Lyapunov exponent, i.e. localization, as guaranteed by Furstenberg's theorem for products of random matrices~\cite{furstenberg1963}. Differently, in the focusing case $U_-(\lambda)$ for $\lambda\in\mathbb{R}$ is traceless and anti-Hermitian, i.e. it belongs to $\mathfrak{su}(2)$, so that $\mathcal{T}(\lambda)\in SU(2)$ is unitary. Unitarity forbids any exponential growth of the solutions of \eqref{eq:zs1}, and localization is therefore impossible.

At the level of the TBA, this means that in the defocusing case we expect a purely radiative TBA, in agreement with the semiclassical limit of the repulsive Lieb-Liniger model~\cite{del2020exact}. We expect this can be derived by reproducing the radiative part of the procedure of Sec.~\ref{sec:pf}. In the focusing case, instead, the absence of localization makes radiation subleading in the thermodynamic limit, in agreement with the semiclassical limit of the attractive Lieb-Liniger model~\cite{koch22}. It, then, remains a purely solitonic TBA, but these solitons have a non-standard entropy, different from the expected Maxwell-Boltmann one. The reason is that the solitons of the focusing NLS, labelled by a complex eigenvalue $\lambda_j$, have a spatial extent controlled by $\mathrm{Im}\,\lambda_j$ which is independent of their velocity, so they can be arbitrarily extended. This drastically changes their counting and hence their entropy. 
The same structure is also shared by the Ablowitz-Ladik chain, the integrable discretization of the NLS equation~\cite{Ablowitz2003}.

The sinh- and sine-Gordon equations fit in the same scheme, the only difference being the relation between the potentials and the initial-time field~\cite{Faddeev87}. We will not give further details, and only comment that the two cases mirrors that of the two NLS equations: for the sinh-Gordon model the transition matrix belongs to $SU(1,1)$, as for the defocusing NLS, so that localization holds and the TBA is purely radiative. For the sine-Gordon model it belongs to $SU(2)$, as for the focusing NLS, so that localization is absent, radiation is subleading, and the TBA is purely solitonic. In the latter case the solitonic content consists of kinks and antikinks, whose width is fixed by the mass scale, and of breathers, which, like the NLS solitons, can be arbitrarily extended. It is then the breather sector that carries the non-standard entropic term, consistently with the semiclassical limit of the quantum sine-Gordon TBA~\cite{Koch23}.

At present we lack a technique to extend our procedure to such extended solitons. This will be the subject of future work and constitutes the main limitation of our approach.

\section*{Acknowledgements}
I am indebted to Herbert Spohn for pointing me to this problem and for continuous discussions and support. I also thank Alvise Bastianello and Simone Warzel for useful discussions and comments.
I acknowledge the financial support from the Deutsche Forschungsgemeinschaft (DFG, German Research Foundation) – TRR 352 – Project-ID 470903074.

%==========================================================
%==========================================================
%==========================================================
\bibliographystyle{SciPost_bibstyle}
\bibliography{refs}

%==========================================================
%==========================================================
%==========================================================
\clearpage
\appendix

%==========================================================
%==========================================================
%==========================================================
\section{The periodic problem}
\label{app:periodic}

Consider a finite Toda chain of $N$ sites with periodic boundary conditions, $a_{j+N}=a_j$, $p_{j+N}=p_j$. The spectrum of the periodic Lax matrix is encoded in the \emph{monodromy matrix}, i.e.\ the transfer matrix \eqref{eq:todaPhi} across one period,
\be
M_N(z)=\Phi(\lambda(z),N,0)=\overset{\curvearrowleft}{\prod_{j=0}^{N-1}} U(\lambda(z),j),
\ee
with $U(\lambda,j)$ the one-step transfer matrix \eqref{eq:todaU}. Periodicity implies $\det M_N=1$, so its eigenvalues are $\mu,\mu^{-1}$ with $\mu+\mu^{-1}=\Delta(z)=\operatorname{Tr}M_N(z)$. A periodic eigenfunction, $\psi_{j+N}=\psi_j$, is an eigenvector of $M_N$ with $\mu=1$, so the spectrum is the solution set of
\be\label{eq_disc}
\Delta(z)=2 .
\ee
Since $\Delta$ is a polynomial of degree $N$ in $\lambda$, \eqref{eq_disc} has $N$ roots, counted with multiplicity. They are real, being the eigenvalues of the $N\times N$ real symmetric Lax matrix.

Consider now an infinite chain in the vacuum, expect for a region of $N$ sites that is given by the configuration of the periodic problem above. For $|z|=1$,  its transition matrix \eqref{eq_fulltm} is given by $\mathcal{T}_\mathrm{R}(z)=R(z)^{-1}\Phi_0(z,-N)\,M_N(z)\,R(z)$, or equivalently
\be
R(z)^{-1}M_N(z)\,R(z)=
\begin{pmatrix}
  z^{-N} & 0 \\ 0 & z^{N}
\end{pmatrix}
\mathcal{T}_\mathrm{R}(z).
\ee
The monodromy is thus the transition matrix dressed by the free propagation across the cell. Taking the trace, for $|z|=1$,
\be\label{eq_discscatt}
\Delta(z)=\alpha(z)\,z^{-N}+\bar\alpha(z)\,z^{N}=2 .
\ee
Only $\alpha$ enters, i.e. the transmission coefficient $T=1/\alpha$. In the vacuum, $\alpha=1$ recovers the free quantization $z^N+z^{-N}=2$, i.e.\ $k_n=2\pi n/N$ with $z=e^{ik}$.

For the radiative modes, $z=e^{ik}$, \eqref{eq_discscatt} reads $2|T(e^{ik})|^{-1}\cos\!\big(kN+\arg T(e^{ik})\big)=2$, i.e.
\be\label{eq_bethe}
kN+\arg T(e^{ik})=2\pi n\pm\arccos|T(e^{ik})|,
\qquad n\in\mathbb{Z},
\ee
This is a Bethe-type quantization, in which the bare momenta are dressed by the transmission phase.

\end{document}